\pdfoutput=1

\documentclass[11pt]{article}

\usepackage[final]{acl}

\usepackage{times}
\usepackage{latexsym}
\usepackage{subcaption}
\usepackage{booktabs}
\usepackage{xcolor}
\usepackage{pifont}
\newcommand{\cmark}{\ding{51}}
\usepackage{multirow} 
\definecolor{darkgreen}{rgb}{0.0,0.5,0.0}
\usepackage[T1]{fontenc}

\usepackage[utf8]{inputenc}
\usepackage{tcolorbox}
\tcbuselibrary{raster,skins,breakable} 

\tcbset{
  colframe=black!15,
  colback=white,
  boxrule=0.4pt,
  arc=2mm,
  left=1mm,right=1mm,top=1mm,bottom=1mm
}

\newtcolorbox{qaquestion}{
  enhanced, breakable, colback=white, colframe=black!25, boxrule=0.45pt, arc=2mm
}
\newtcolorbox{qaanswer}{
  enhanced, breakable, colback=black!3, colframe=black!15, boxrule=0.4pt, arc=2mm
}

\newenvironment{casecard}[1][]%
{%
  \begin{tcolorbox}[enhanced, breakable, colback=black!2, colframe=black!10, boxrule=0.3pt,
    title={\ifx\relax#1\relax\else\textbf{#1}\fi}]
}
{%
  \end{tcolorbox}
}

\usepackage{microtype}
\usepackage{amsmath}
\usepackage{inconsolata}

\title{Speech-Hands: A Self-Reflection Voice Agentic Approach to \\ Speech Recognition and Audio Reasoning with Omni Perception}

\author{
Zhen Wan$^{1,2}$  \hspace{1em}
Chao-Han Huck Yang$^1$ \hspace{1em} 
Jinchuan Tian$^{1,3}$ \hspace{1em} 
Hanrong Ye$^1$ \hspace{1em} 
Ankita Pasad$^1$ \hspace{1em} 
\\
{\bf Szu-wei Fu$^1$ }\hspace{1em} 
{\bf Arushi Goel$^1$ }\hspace{1em} 
{\bf Ryo Hachiuma$^1$} \hspace{1em}
{\bf Shizhe Diao$^1$} \hspace{1em} 
{\bf Kunal Dhawan$^1$} \hspace{1em} 
\\
{\bf Sreyan Ghosh$^1$ }\hspace{1em} 
{\bf Yusuke Hirota$^1$} \hspace{1em} 
{\bf Zhehuai Chen$^1$ }\hspace{1em} 
{\bf Rafael Valle$^1$ }\hspace{1em}
 \\
{\bf Chenhui Chu$^2$ }\hspace{1em}
{\bf Shinji Watanabe$^3$ }\hspace{1em}
{\bf Boris Ginsburg$^1$ }\hspace{1em}
{\bf Yu-Chiang Frank Wang$^1$ }\hspace{1em}
\\
$^1$NVIDIA \hspace{1em}
$^2$Kyoto University \hspace{1em}
$^3$Carnegie Mellon University \hspace{1em}
\\
Corresponding authors:
\texttt{zhenwan@nlp.ist.i.kyoto-u.ac.jp; hucky@nvidia.com} \\
}

\begin{document}
\maketitle
\begin{abstract}
We introduce a voice-agentic framework that learns one critical omni-understanding skill: knowing when to trust itself versus when to consult external audio perception. Our work is motivated by a crucial yet counterintuitive finding: naively fine-tuning an omni-model on both speech recognition and external sound understanding tasks often degrades performance, as the model can be easily misled by noisy hypotheses. To address this, our framework, Speech-Hands, recasts the problem as an explicit self-reflection decision. This learnable reflection primitive proves effective in preventing the model from being derailed by flawed external candidates. We show that this agentic action mechanism generalizes naturally from speech recognition to complex, multiple-choice audio reasoning. Across the OpenASR leaderboard, Speech-Hands consistently outperforms strong baselines by 12.1\% WER on seven benchmarks. The model also achieves 77.37\% accuracy and high F1 on audio QA decisions, showing robust generalization and reliability across diverse audio question answering datasets. By unifying perception and decision-making, our work offers a practical path toward more reliable and resilient audio intelligence. \footnote{Project page, interactive demo, openclaw branch, and code: \url{https://YukinoWan.github.io/Speech-Hands/}}

\end{abstract}

\section{Introduction}

Omni-modal models~\cite{xie2024miniomnilanguagemodelshear,openai2024gpt4technicalreport,xu2025qwen25omnitechnicalreport,li2025baichuanomni15technicalreport} that jointly process audio and text have unified a range of audio understanding tasks, including automatic speech recognition (ASR), temporal sound event reasoning, and knowledge-heavy question answering. However, human perception is not naturally perfect at understanding acoustic patterns across different resolutions at soundscape~\cite{Calcus2024}. For example, while professional speech interpreters could produce superior ASR transcriptions, this specialized ability does not guarantee a comparable aptitude for understanding animal sounds or complex music~\cite{Galantucci2006}.

\begin{figure}[t]
    \centering
    \includegraphics[width=0.9\linewidth]{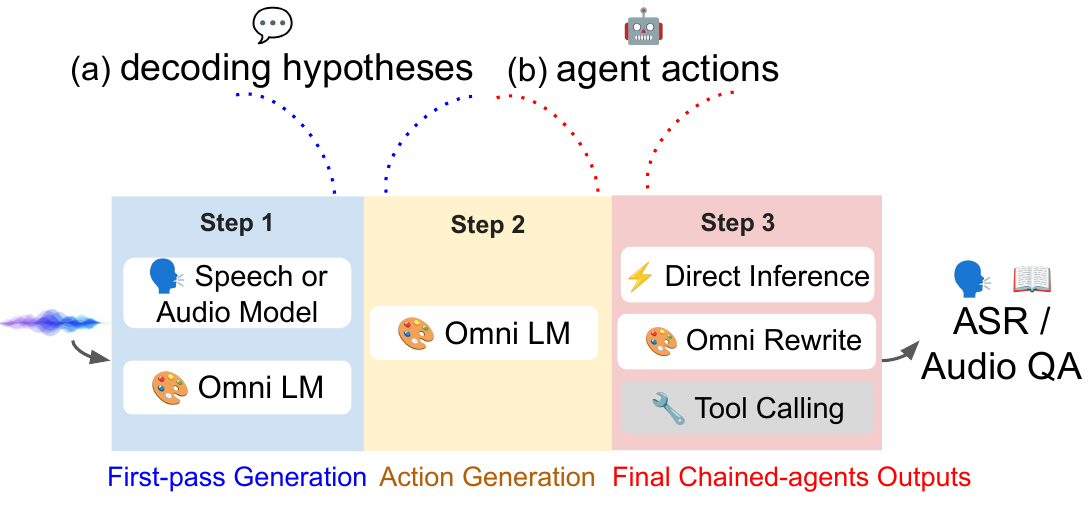}
    \caption{\textit{Speech-Hands} acts as a dynamic orchestrator that predicts a special action token to govern its cognitive strategy for ASR and multi-domain audio reasoning.}
    \label{fig:overview}
\end{figure}

Inspired by developmental psychology~\citep{selman1974structural}, we draw a parallel to the human capacity for self-reflection, where children mature from a purely egocentric viewpoint to a stage of \textit{self-reflective perspective-taking}, 
which serves the critical ability to ``step outside'' one's own thoughts, evaluate one's beliefs against others', and importantly, to recognize the boundaries of one's own knowledge. In contrast, current models often operate egocentrically, implicitly trusting their internal perception without the capacity to critically assess its reliability or seek external assistance when necessary. We aim to instill a form of computational self-reflection~\cite{Nelson1990} into an omni-modal agent, designing a collaborative framework that explicitly reasons about when to trust its own perception, when to defer to an expert, and even when to utilize tools.

We frame these voice understanding models not just as passive predictors, but as agentic that have access to multiple internal and external information sources, and must decide how to best use them. For such an agent, a central decision-making challenge arises~\cite{Lebiere2011}: should it rely on its own auditory perception, or consult external suggestions, such as ASR alternatives or other perceptual consultants? Prior work, such as ASR and large language model (LLM) cascaded approach of Generative Error Correction (GER)~\citep{yang2023generative,lin2025neko}, as shown in Figure~\ref{fig:extension}, sidesteps this agentic dilemma entirely. By operating only on text hypotheses without access to the original audio, these methods are fundamentally non-agentic; they cannot weigh internal perception against external advice because they have no internal perception to begin with. Our preliminary experiments reveal that naively combining modalities often degrades performance, as the model struggles to resolve conflicts between its own perception and flawed external suggestions~\cite{Kaiser2021}. Without a mechanism to decide which source to trust, the model is easily confused. 

To address this, we introduce \textit{Speech-Hands}, a learnable framework that instantiates self-reflection as a core control primitive. 
As illustrated in Figure~\ref{fig:overview}, our agent operates as a dynamic orchestrator. It begins by aggregating multi-source decoding hypotheses (a) from the first-pass generation.
Instead of blindly fusing inputs, the agent critically evaluates them to predict an explicit agent action (b) during the action generation phase.
This control token effectively dicates the model's cognitive strategy: trigerring fast direct inference when confidence is high (selecting whether its internal perception or external perceptions), engaging in omni rewrite over available evidence, or initiating tool calling when special utilities are required. In this work, we will focus on the actions of direct inference and omni rewrite, leaving the tool calling action in future work.
This approach unifies transcription and reasoning under a single, controllable framework that knows when to trust, when to rethink, and when to ask for help. 
\color{black}{}

\color{black}{
\subsection{Preliminary: The Surprising Failure of Multimodal Correction with Omni-LM}

A natural hypothesis is that providing an omni model with both audio and text hypotheses during supervised fine-tuning (SFT) should enhance GER performance. We tested this assumption by fine-tuning Qwen2.5-Omni~\cite{xu2025qwen2} to correct N-best hypotheses ($N=5$) from Whisper-v2-large~\cite{radford2023robust} on OpenASR datasets.

\begin{figure}[ht!]
    \centering
    \includegraphics[width=0.98\linewidth]{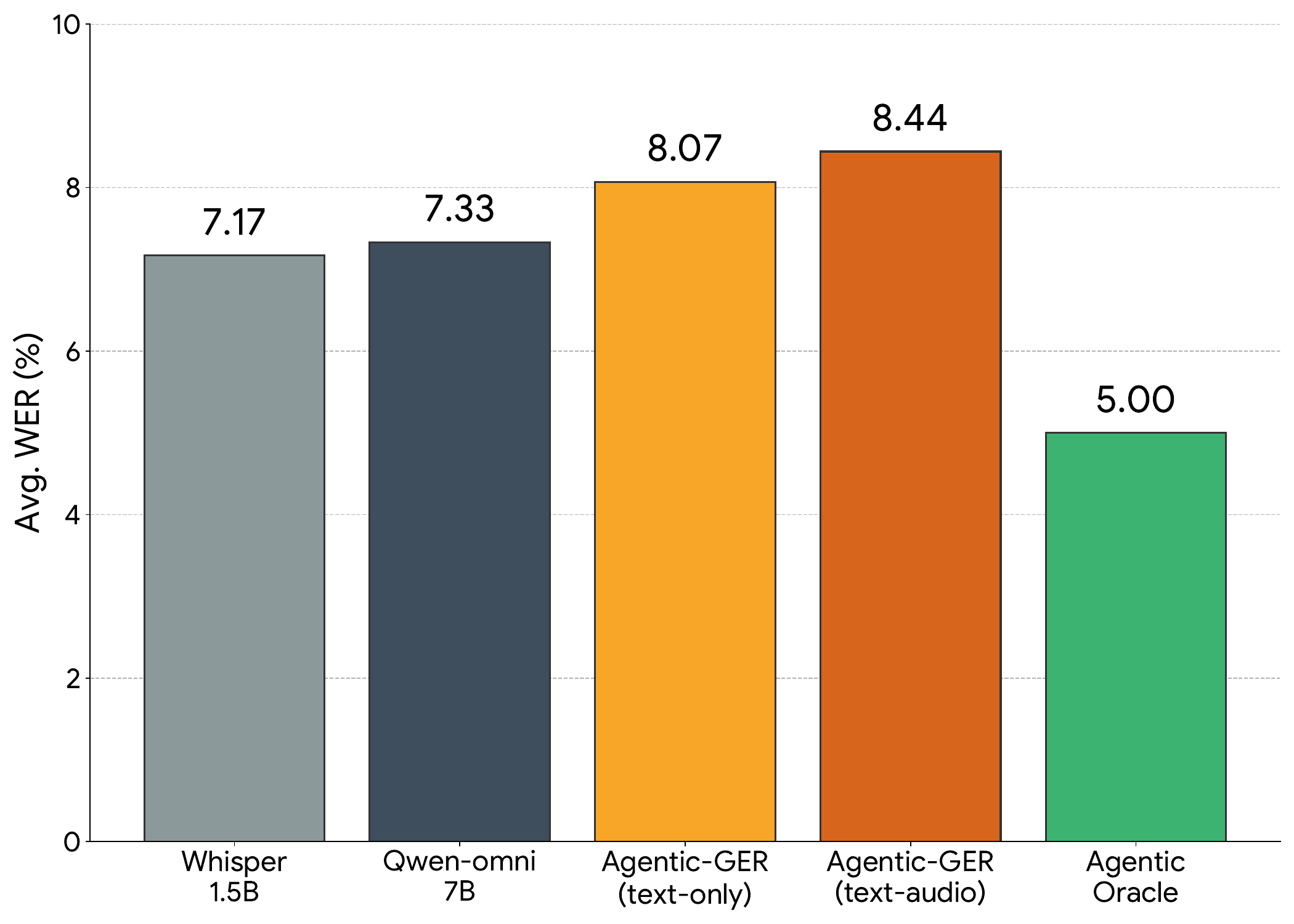}
    \caption{Preliminary results on the cascaded agentic Qwen-omni baseline for generative error correction (GER) with supervised fine-tuning show that both text-only and text-audio GER degrade ASR performance, where the best ASR and LLM combination achieves a low agentic output oracle of 5\% WER.}
    \label{fig:2:wer}
\end{figure}

Figure~\ref{fig:2:wer} results, however, are surprisingly negative. As shown in our preliminary study (Table~\ref{tab:wer_results}), this naive SFT approach consistently degrades performance across seven ASR benchmarks, yielding a higher Word Error Rate (WER) than either of the baseline models alone. }

To verify that this degradation was not due to suboptimal prompting, we extensively tested varying instructions during SFT, which aims to emphasize internal audio perception, external transcripts, or a balanced fusion.
However, as shown in Table~\ref{result:prompt_ablation}, all prompting setups failed to recover performance (e.g., WER increased to 8.52\%--9.05\% compared to baselines).

\begin{table}[h!]
\centering
\small
\setlength{\tabcolsep}{6pt}
\begin{tabular}{lrrr}
\toprule
\textbf{AVG. WER} & (a) & (b)\\
\midrule
Emphasize internal & - & 8.63\\
Emphasize external   &8.58 & 8.67\\
Emphasize audio     & 9.02 & 9.05\\
Balanced     & 8.44 & 8.52\\
\bottomrule
\end{tabular}
\caption{Prompt ablation results in preliminary SFT. (a) audio + external whisper 5-best and (b) audio + internal 1-best + external whisper 5-best. Details in Appendix~\ref{app:prompt_ablation}}
\label{result:prompt_ablation}
\end{table}

\begin{figure*}[t]
    \centering
    \includegraphics[width=0.8\linewidth]{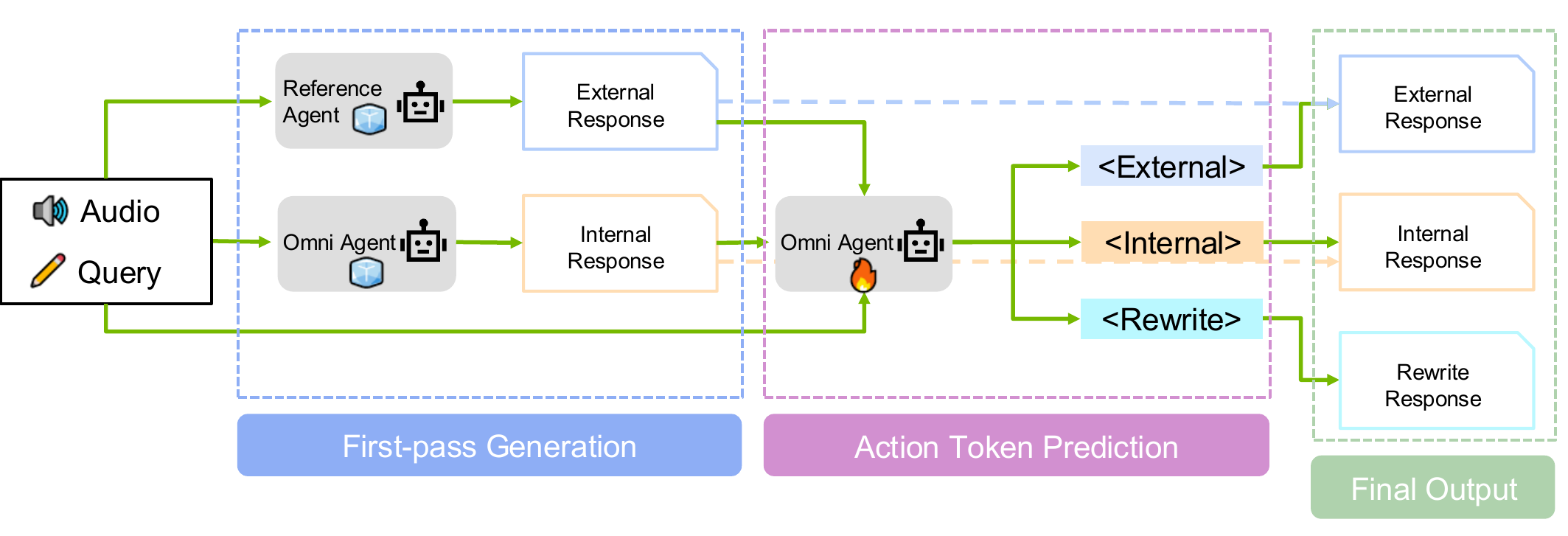}
    \caption{Overview of our proposed Self-Reflection Multimodal GER framework. A special token is generated at the beginning to decide whether to use audio perception (i.e., transcription hypotheses or caption) or not.}
    \label{fig:selfrag_overview}
\end{figure*}

Furthermore, our zero-shot analysis reveals that the base model lacks intrinsic arbitration capabilities: its decisions are highly sensitive to prompt wording rather than ground truth, often collapsing into trivial heuristics (as shown in Table \ref{tab:zeroshot_matrix} and details in Appendix~\ref{app:zeroshot_analysis}).

\begin{table}[h]
    \centering
    \caption{Confusion matrix of zero-shot decisions under different prompting strategies. The results show that the model's arbitration is highly sensitive to prompt wording rather than the ground truth correctness (Oracle), often collapsing into trivial heuristics.}
    \label{tab:zeroshot_matrix}
    \resizebox{\linewidth}{!}{
    \begin{tabular}{l|rr|rr|rr}
        \toprule
        \multirow{2}{*}{\textbf{Ground Truth}} & \multicolumn{2}{c|}{\textbf{Internal-biased Prompt}} & \multicolumn{2}{c|}{\textbf{External-biased Prompt}} & \multicolumn{2}{c}{\textbf{Balanced Prompt}} \\
        \cmidrule(lr){2-3} \cmidrule(lr){4-5} \cmidrule(lr){6-7}
         & Pred. Int & Pred. Ext & Pred. Int & Pred. Ext & Pred. Int & Pred. Ext \\
        \midrule
        \textbf{Oracle Internal} & 0.83 & 0.17 & 0.35 & 0.65 & 0.68 & 0.32 \\
        \textbf{Oracle External} & 0.71 & 0.29 & 0.14 & 0.86 & 0.34 & 0.66 \\
        \bottomrule
    \end{tabular}
    }
\end{table}

These findings demonstrate a fundamental flaw in the naive omni-LM fusion approach: without a mechanism to adjudicate between its own perception and potentially flawed external advice, the omni-model is easily confused, often amplifying hallucinations or overcorrections as shown in the ASR failure case (Appendix~\ref{case_ger_fail} and Section~\ref{cases}). This provides strong motivation for a more principled mechanism that allows the model to learn when and how to incorporate external information.

\section{Related Work}

\subsection{Voice Retrieval and Agentic Framework}
Voice retrieval has become a useful component in augmenting audio tasks such as captioning~\citep{koizumi2020audiocaptioningusingpretrained, zhao-etal-2023-generating}, audio-to-text generation~\citep{huang2023makeanaudiotexttoaudiogenerationpromptenhanced}, dialogue system~\cite{chen2025wavragaudiointegratedretrievalaugmented}, and music generation~\citep{gonzales-rudzicz-2024-retrieval}. Recently, voice retrieval modules have been incorporated into multimodal agents~\citep{yang2023mmreactpromptingchatgptmultimodal, zhang2024omagentmultimodalagentframework,wang2025audioagentleveragingllmsaudio, wan2025speechiq}, providing access to external memory or specialized tools across modalities.

Yet, even in these agentic frameworks, retrieval is often treated as an auxiliary enhancer, rather than as a distinct source of information. When the retrieved content diverges from the internal prediction of the model, current systems lack principled mechanisms for arbitration.

\color{black}{}
\subsection{Omni-Modality and Self-Reflection in Multimodal Models}

By weaving together text, vision, and audio into a single fabric of understanding, these systems begin to approach the fluidity of human perception~\cite{xu2025qwen2, xu2025qwen3, goel2024omcat, abouelenin2025phi}.  Modality-specific reflection methods~\cite{hu-etal-2025-investigating} suggest that introspection within each sensory channel can partially bridge these gaps, aligning representations with quiet precision.

Recent works also introduce explicit self-reflection~\cite{renze2024self, madaan2023self} into multimodal reasoning~\cite{cheng2024vision, fang2025fewer}. The self-reflected video reasoner~\cite{songmodularized} iteratively critiques its own visual understanding to reinforce policy stability, while other efforts call for reflective checks against overconfidence and modality neglect~\cite{yang2025humanomniv2}. Still, these mechanisms operate \emph{after} perception, which treats reflection as a corrective mirror once fusion has already occurred.

Rather than reflecting on the output, our Speech-Hands framework reflects on the act of perception itself. It learns an action mechanism that decides, in real time, whether to trust its own \textit{``ears''} or the \textit{``words''} of others. We aim to discover self-reflection from a post-hoc repair strategy into a preemptive act of perceptual discernment toward an early glimmer of meta-cognition within multimodal understanding.
\color{black}{}

\color{black}{}
\section{Methodology}
\label{sec:methodology}

We present \textbf{Speech-Hands}, a learnable omni-agentic framework for audio understanding and reasoning (Figure~\ref{fig:selfrag_overview}). It enables a multimodal language model to explicitly choose a special token: trusting its own internal perception or defer to external hypotheses, enables efficient \textbf{Fast Inference}, while rewriting a new response engages the \textbf{Omni Rewrite} process for deeper reasoning. This token is generated during inference and guides the downstream generation process, allowing interpretable and agentic decision-making.

\subsection{Task Formulation}
\label{sec:task_formulation}

We formulate a unified self-reflection agent that generalizes across both speech recognition and audio question answering. Given an input audio $A$, an optional query $Q$, the agent first generates its own response $H_\text{omni}$, and then combined with an external response $H_\text{ext}$ provided by an external model. 

Next, rather than fusing these sources implicitly as normal GER researches, Speech-Hands introduces a learned policy to explicitly choose among them. Our agent model first emits a special action token from the set $\{\texttt{<internal>}, \texttt{<external>}, \texttt{<rewrite>}\}$ to indicate whether to trust itself or rely on external hypothesis or even whether rewriting a new response after rethinking the audio task and all input resources for the final answer (GER). This self-reflection decision is made based on the full context $(A, Q, H_\text{omni}, H_\text{ext})$, and the selected action conditions the final generation.



To supervise the self-reflection mechanism, we construct ground-truth action token labels on the whole training dataset by comparing the performance of internal, external, and GER predictions. We leverage detailed strategies for action token construction in ASR and audio QA.
\subsection{Action Token Construction for ASR}

 In the ASR setting, we use WER as a pointer to decide our actions. For each audio example, we first prompt the omni model to generate the transcript $T_{\text{int}}$ and in parallel leverage an ASR model to predict the external $T_{\text{ext}}$, we then let omni model to generate again based on both the audio and the external $T_{\text{ext}}$ to acquire the GER prediction $T_{\text{ger}}$. Next, we compute the WER between the ground-truth transcript $T_{\text{gt}}$ and each of the three candidates. If $T_{\text{int}}$ is the same as $T_{\text{gt}}$ ($WER=0$) or has the lowest WER, the label is assigned as \texttt{<internal>}, this is to encourage the model to trust itself when it can successfully solve the problem, otherwise \texttt{<external>} if $T_{\text{ext}}$ performs best, or \texttt{<rewrite>} if $T_{\text{ger}}$ performs best.

\subsection{Action Token Construction for Audio QA}

Unlike the ASR setting where token selection is based on fine-grained WER scores, Audio QA presents a discrete supervision signal: each prediction is either correct or incorrect. For each instance consisting of an audio segment $A$, a question $Q$, and answer choices $\mathcal{C}$, we first prompt the omni model to produce an internal prediction $c_{\text{int}}$ based on $(A, Q)$. In parallel, we obtain an external prediction $c_{\text{ext}}$ from an audio reasoning model.

We then compare both predictions against the ground-truth answer $c^*$. If $c_{\text{int}} = c^*$, we assign the label \texttt{<internal>}, encouraging self-reliance when the model performs well. If $c_{\text{int}}$ fails but $c_{\text{ext}} = c^*$, we assign \texttt{<external>} to delegate control. Otherwise, when both predictions are incorrect, the label is set to \texttt{<rewrite>}, signaling a need to re-evaluate the question with all available context.

However, this binary decision process introduces instability during training. External predictions can be stochastic. Therefore, repeated sampling may yield different answers, especially under high uncertainty or directly change another external model can also lead to a different accuracy. This stochasticity makes the decision boundary between \texttt{<external>} and \texttt{<rewrite>} inherently less robust.

To mitigate this, we adopt a multiple decoding-based strategy: for each example, we sample the external model five times and collect their predicted choices. If the majority of predictions match the ground-truth answer, we assign \texttt{<external>}; otherwise, we assign \texttt{<rewrite>}. This approach stabilizes supervision by reducing the variance in external outputs and yields more reliable action labels.

\subsection{Prompt Formatting and Training}

Subsequently, each training instance is formatted as a single target string consisting of the decision token followed by the final target transcript or answer, e.g. \texttt{<rewrite>} + ground-truth transcription. This unified string allows the model to learn not only how to generate the task output but also how to decide the action before generation.
We adopt an instruction-style prompt to guide the model to make decisions, below is the prompt template for ASR:

\begin{quote}
\small
\ttfamily
You are an omni-agent for speech understanding with access to three inputs:\\
(1) The original audio;\\
(2) Five transcription hypotheses from another ASR system (external);\\
(3) Your own first-pass transcription (internal).\\

Your task is to:\\
- First decide whether your internal transcription is reliable.\\
- If yes, output <internal> and your transcription.\\
- If the external system is more reliable, output <external> and use one of its hypotheses.\\
- Otherwise, output <rewrite> and generate a new answer using both sources and the audio.
\end{quote}
For Audio QA, the prompt tempalte is shown in Appendix ~\ref{audioqa_prompt}.  
During training, we optimize a single cross-entropy loss over the concatenated target sequence, which jointly supervises the action token and the subsequent prediction. Concretely, the model first predicts the action token and then continues decoding the target transcript or answer; both parts contribute to the same loss. This end-to-end objective encourages the model to internalize the mapping from multimodal evidence to action choice and to generate the corresponding output under that decision.

\begin{table*}[t]
\centering
\small
\setlength{\tabcolsep}{4.3pt}
\resizebox{1.0\linewidth}{!}{
\begin{tabular}{lrrrrrrr|r}
\toprule
\textbf{Dataset} & \textbf{AMI} & \textbf{Tedlium} & \textbf{Gigspeech} & \textbf{Spgispeech} & \textbf{VoxPopuli} & \textbf{Libri-clean} & \textbf{Libri-other} & \textbf{avg. WER $\downarrow$} \\
\midrule
\rowcolor{blue!10} \multicolumn{9}{l}{\textbf{ASR model or Omni-LLM}} \\
Whisper-v2-large     & 16.88 & 4.32 & 11.45 & 3.94 & 7.57 & 2.91 & 5.15 & 7.17 \\
Canary-1b-v2         & 19.80 & 4.78 & 11.66 & 3.08 & 6.35 & 1.73 & 3.17 & 7.22 \\
Parakeet-tdt-0.6b-v3 & 12.69 & 4.90 & 12.24 & 3.16 & 6.48 & 1.89 & 3.37 & 6.68 \\
\midrule
Qwen2.5\_omni        & 19.77 & 5.17 & 11.26 & 4.58 & 6.59 & 2.09 & 3.85 & 7.33 \\
Phi-4-MM   & 11.69 & \textbf{2.90} & \textbf{9.78} & 3.13 & \textbf{5.93} & 1.68 & 3.83 & 6.14 \\
Gemini-2-Flash & 21.58 & 3.01 & 10.71 & 3.82 & 7.89 & 2.49 & 5.84 & 8.56 \\
GPT-4o-voice   & 57.76 & 5.79 & 13.64 & 5.66 & 10.83 & 3.48 & 7.97 & 15.76 \\
\midrule
\rowcolor{orange!10} \multicolumn{9}{l}{\textbf{Qwen2.5-Omni: Cascaded}} \\
GER: $\Rightarrow$ Whisper-v2-large       & 23.44 & 6.15 & 12.15 & 3.94 & 7.53 & 2.97 & 4.89 & 8.44 \\
GER: $\Rightarrow$ canary       & 24.58 & 6.38 & 12.43 & 4.02 & 7.72 & 3.05 & 5.01 & 8.74 \\
GER: $\Rightarrow$ parakeet      & 22.91 & 6.09 & 12.10 & 3.98 & 7.49 & 2.92 & 4.84 & 8.33 \\
\midrule
\rowcolor{green!15} \multicolumn{9}{l}{\textbf{Qwen2.5-Omni: Parallel}} \\
Speech-Hands $\rightleftharpoons$ whisper-v2      &15.03 & 4.45 & 12.37 & 3.01 & 6.49 & 1.86 & 3.46 & 6.67 \\
Speech-Hands $\rightleftharpoons$ canary       & 15.29 & 4.21 & 10.87 & \textbf{2.17} & 5.96 & \textbf{1.61} & \textbf{3.07} & 6.17 \\
Speech-Hands $\rightleftharpoons$ parakeet     & \textbf{11.20} & 4.37 & 11.10 & 2.26 & 6.02 & 1.67 & 3.18 & \textbf{5.69} \\
\bottomrule
\end{tabular}
}
\caption{WER (\%) results across 7 datasets, with the average WER shown in the rightmost column. Speech-Hands training significantly outperforms both baseline systems (ASR model, Qwen) and prior cascaded GER setups.}
\label{tab:wer_results}
\end{table*}

\subsection{Agentic Inference via Action Tokens}

At inference time, the model receives the same multimodal inputs as during training. 
The model performs a two-stage decoding process: it first emits an action token: \texttt{<internal>}, \texttt{<external>}, or \texttt{<rewrite>}. The decoding process is subject to decide which information source to prioritize, and then generates the final output accordingly.
This explicit agentic self-reflection mechanism offers interpretability and control over how the model balances internal perception and external knowledge. It enables direct analysis (\textit{e.g.}, F1 score) of when the model relies on its own understanding, consultants from external systems, or synthesizes a new response. The unified prediction format ensures that the model not only learns what to generate, but also which to trust across both speech recognition and audio reasoning tasks.


\section{Experimental Setup}

\subsection{Datasets}

\textbf{Speech Recognition.} 
We use seven representative datasets covering a range of domains, styles, and noise conditions from OpenASR leaderboard: AMI~\cite{7280661f53654898abc5962f926ba81a} (meeting speech), Tedlium~\cite{Hernandez_2018} (TED talks), GigaSpeech~\cite{Chen_2021} (large-scale podcasts and YouTube-style speech), SPGISpeech~\cite{oneill21_interspeech} (long-form read speech), VoxPopuli~\cite{wang-etal-2021-voxpopuli} (English subset of multilingual political recordings), and LibriSpeech~\cite{7178964} (clean and noisy audiobook speech). For baseline fine-tuning and prompt-based GER, we use all available training sets. For our proposed method, unless otherwise specified ( \texttt{w/ FULL Datasets}), we train on at most 20,000 examples per dataset, this is due to the limitation of heavy computation requirements when doing inference on the whole training set for internal, external and GER (token distribution is discussed in ~\ref{sec:special_token_analysis}).

\textbf{Audio Reasoning.} We evaluate on the multi-domain audio question-answering benchmark~\cite{yang2025multidomainaudioquestionanswering} (MD-Audio), which consists of multiple-choice questions grounded in real audio clips. This benchmark includes three complementary subsets that probe different reasoning capabilities. While previous audio reasoning benchmarks such as MMAU~\cite{sakshi2024mmau} and MMAR~\cite{ma2025mmar} provide only MMLU-style test sets, MD-Audio releases both training and development sets, enabling evaluation of the proposed trainable agentic framework, as shown in Table~\ref{tab:audioqa_stats}. Detailed dataset descriptions can be found in Appendix~\ref{data_detail}. For each sample, we construct inputs as described in Section~\ref{sec:methodology}, including the original audio, a first-pass internal prediction from Omni model, one external hypothesis and GER result (for ASR).

\section{Results}
\begin{table*}[t]
\centering
\small
\setlength{\tabcolsep}{4.5pt}
\begin{tabular}{lrrr|r}
\toprule
\textbf{Model / Setting} & \small{\textbf{Bio-acoustic}} & \small{\textbf{Soundscape}} & \small{\textbf{Complex QA}} & \textbf{avg. Acc. $\uparrow$} \\
\midrule
\rowcolor{blue!10} \multicolumn{5}{l}{\textbf{Audio LM or Omni Model}} \\
Gemini-2-Flash  & 42.03 & 46.34 & 59.89 & 56.61 \\
Qwen2.5-Omni  & 47.32 & 56.32 & 59.89 & 57.87 \\
AudioFlamingo 3 (AF3) & 71.88 & 57.31 & 81.26 & 74.49 \\
\midrule
\rowcolor{gray!10} \multicolumn{5}{l}{\textbf{Qwen2.5-Omni Baselines}} \\
+ SFT with official training data & 78.13 & 34.65 & 76.61 & 63.13 \\
+ GRPO with official training data & 78.09 & 39.43 & 79.12 & 65.54 \\
\rowcolor{gray!20} + GRPO with external audio data  ~\cite{Li_mlpxc_2025} & 62.32 & \textbf{72.10} & 82.15 & 75.10 \\
GER: $\Rightarrow$ AF3 (cascaded agentic) & 76.29 & 52.02 & 77.48 & 68.93 \\
\midrule
\rowcolor{green!10} \multicolumn{5}{l}{\textbf{Qwen2.5-Omni: Speech-Hands }} \\
$\rightleftharpoons$ (parallel agentic): SFT with official training data   & 67.86 & 58.29 & 83.34 & 75.75 \\
$\rightleftharpoons$ (parallel agentic) + majority sampling & \textbf{81.25} & 59.4 & \textbf{85.7} & \textbf{77.37} \\
\bottomrule
\end{tabular}
\caption{AudioQA and acoustic content reasoning accuracy (\%) across knowledge-intensive bioacoustic QA~\cite{sayigh2016watkins}, multi-sound-object soundscapes, and MMAU-style~\cite{sakshi2024mmau} complex audio QA tasks. }
\label{tab:audioqa_results}
\end{table*}

\subsection{Training Details}

All experiments are conducted using the Qwen2.5-Omni model. We extend its tokenizer to include three special action tokens, \texttt{<internal>}, \texttt{<external>} and \texttt{<rewrite>}, used during both training and inference. Models are trained using the standard supervised fine-tuning (SFT) objective with a cross-entropy loss.
We train for 5 epochs with fp16. The batch size is set to 64, and the learning rate is initialized at 1e-4 with cosine decay. All experiments adopt greedy decoding.

\subsection{Baselines}

\textbf{External ASR models}: We include three high-performing supervised speech recognition models as external references: Whisper-v2-large~\cite{radford2022robustspeechrecognitionlargescale}, Canary-1B-v2, and Parakeet-TDT-0.6B-v3~\cite{sekoyan2025canary1bv2parakeettdt06bv3efficient}. These systems operate in a closed transcription setting and provide non-generative references.

\textbf{External audio QA model}: For multi-choice audio QA, we include Audio Flamingo 3~\cite{goel2025audioflamingo3advancing}, a speech-language model with audio frontend capabilities serving as a strong baseline.

We also include the latest model in comparison, e.g., Phi-4-MM~\cite{microsoft2025phi4minitechnicalreportcompact}, Gemini-2-Flash, GPt-4o-voice in ASR evaluations.

\begin{table*}[t]
\centering
\small
\setlength{\tabcolsep}{4pt}
\resizebox{0.9\linewidth}{!}{
\begin{tabular}{lrrrrrrr}
\toprule
\textbf{Dataset} & \textbf{AMI} & \textbf{Tedlium} & \textbf{Gigaspeech} & \textbf{SPGIspeech} & \textbf{Voxpopuli} & \textbf{Libri-clean} & \textbf{Libri-other} \\
\midrule
\rowcolor{green!10} \multicolumn{8}{l}{\textbf{Training Distribution}} \\
\texttt{<internal>}   & 67.95\% & 86.48\% & 87.76\% & 96.25\% & 93.73\% & 98.96\% & 98.96\% \\
\texttt{<external>} & 31.01\% & 11.18\% & 11.8\% & 3.64\%& 6.09\% & 0.96\% & 0.96\% \\
\texttt{<rewrite>} & 1.04\% & 2.34\% & 0.44\% & 1.21\% & 0.18\% & 0.1\% & 0.1\%  \\
\midrule
\rowcolor{green!10} \multicolumn{8}{l}{\textbf{Test Distribution}} \\
\texttt{<internal>}   &70.28\% & 83.57\% & 85.94\% & 95.42\% & 92.68\% & 98.92\% & 98.75\% \\
\texttt{<external>} &26.91\% & 15.12\% & 12.41\% & 3.81\%& 6.47\% & 0.98\% & 1.08\% \\
\texttt{<rewrite>} & 2.27\% & 1.31\% & 1.65\% & 0.77\% & 0.85\% & 0.1\% & 0.17\%  \\
\midrule
\rowcolor{blue!10} \multicolumn{8}{l}{\textbf{\texttt{<internal>} on Test}} \\
Precision & 0.85 & 0.63 & 0.77 & 0.89 & 0.85 & 0.88 & 0.83 \\
Recall    & 0.78 & 0.72 & 0.90 & 0.94 & 0.90 & 0.99 & 0.99 \\
F1        & 0.81 & 0.67 & 0.83 & 0.91 & 0.87 & 0.94 & 0.9 \\
\midrule
\rowcolor{blue!10} \multicolumn{8}{l}{\textbf{\texttt{<external>} on Test}} \\
Precision & 0.88 & 0.96 & 0.83 & 0.82 & 0.71 & 0.81 & 0.72 \\
Recall    & 0.89 & 0.81 & 0.78 & 0.76 & 0.60 & 0.73 & 0.75 \\
F1        & 0.89 & 0.88 & 0.80 & 0.79 & 0.65 & 0.77 & 0.74 \\
\midrule
\rowcolor{blue!10} \multicolumn{8}{l}{\textbf{\texttt{<rewrite>} on Test}} \\
Precision & 0.62 & 0.33 & 0.32 & 0.52 & 0.50 & 0.0 & 0.0 \\
Recall    & 0.24 & 0.21 & 0.05 & 0.36 & 0.21 & 0.0 & 0.0 \\
F1        & 0.39 & 0.28 & 0.08 & 0.43 & 0.33 & 0.0 & 0.0 \\
\bottomrule
\end{tabular}
}
\caption{Training distribution and test-time F1 scores for Speech-Hands' action tokens.}
\label{tab:special_token_f1}
\end{table*}

\subsection{ASR Results}

Table~\ref{tab:wer_results} presents WER results across seven diverse datasets, and we find that:
(1) Speech-Hands outperforms all baselines. Furthermore, our approach with strong ASR models (canary and parakeet) achieves the lowest WER, even with only 20k training examples, outperforming both ASR models or Omni-LLMs; 
(2) The promtp GER over Whisper lags significantly behind token-based methods. This underscores the importance of explicit control via action tokens rather than relying solely on natural-language prompts;
(3) While pre-trained models like Whisper and Qwen perform well on curated datasets such as LibriSpeech-clean, their performance degrades significantly on conversational benchmarks like AMI. Notably, despite Qwen's relatively weaker base ASR performance, our framework enhances its generalization to the extent that it surpasses stronger baselines such as Phi-4-MM on the average performance, demonstrating the stability and transferability of Speech-Hands on both clean and noisy datasets.

\subsection{AudioQA Results}
\label{sec:audioqa_results}
Table~\ref{tab:audioqa_results} reports accuracy on each sub-task and overall average accuracy:
(1) Our final setup (Speech-Hands + majority sampling) achieves the highest average accuracy (77.37\%), outperforming all baselines and pre-trained models. It performs particularly well on Complex QA (85.70\%) and Bioacoustics QA (81.25\%), indicating its ability to handle both abstract reasoning and fine-grained audio patterns;
(2) Standard supervised fine-tuning (SFT) and prompt-based GER exhibit mixed results. SFT achieves good accuracy on Bioacoustics (78.13\%) but fails on Soundscapes (34.65\%). Prompt-based GER also fails on both Soundscape and Complex QA compared with flamingo 3 baseline.
These results highlight the robustness of our agentic framework in diverse audio reasoning settings.

\section{Additional Analysis}
\subsection{Accuracy of Action Token Prediction}
\label{sec:special_token_analysis}

We analyze the model’s ability to correctly emit the three action tokens of \texttt{<internal>}, \texttt{<external>}, and \texttt{<rewrite>}, in the ASR setting under the Action Tokens + Whisper configuration. 
These tokens interpret whether the model is making correct decisions to guide its final prediction.

Table~\ref{tab:special_token_f1} shows both the training distribution and the test-time precision, recall, and F1 scores. The distribution highlights a strong internal bias: across all datasets, \texttt{<internal>} dominates, exceeding 95\% in Libri-clean, Libri-other, spgispeech, and Voxpopuli. In contrast, \texttt{<external>} is sparsely supervised, where below 1\% in Librispeech and \texttt{<rewrite>} is extremely rare everywhere, often less than 2\%. This imbalance poses a natural challenge for learning reliable action.

Despite this skew, the model demonstrates robust performance for the two tokens. On test data, \texttt{<internal>} predictions achieve F1 scores above $0.8$ on most datasets ($0.91$ on spgispeech, $0.94$ on Libri-clean, $0.90$ on Libri-other), indicating that the model can reliably recognize when its own decoding is sufficient. Even for the much rarer \texttt{<external>} token, the model attains high F1 scores, showing strong generalization of deferring to external hypotheses despite limited supervision.

The \texttt{<rewrite>} token proves to be the most challenging, with F1 scores below 0.4 in all but one dataset and zero in Librispeech, where positive training examples are extremely rare. A closer examination reveals that precision consistently exceeds recall, indicating that when the model does emit \texttt{<rewrite>}, its decision is generally correct but under-triggered in the omni model. This suggests a cautious yet reasonably reliable rewrite detector, whose coverage could be further improved through targeted data augmentation. 

Overall, these results validate the effectiveness of the proposed agentic action: even under heavy class imbalance, the model learns to accurately identify when to trust its own predictions versus when to consult external information. The main bottleneck remains the \texttt{<rewrite>} case, suggesting that richer sampling or augmentation strategies may be needed to stabilize this decision in future work.
\color{black}{}

\begin{figure*}[t]
  \centering
  \begin{subfigure}[t]{0.32\linewidth}
    \centering
    \includegraphics[width=\linewidth]{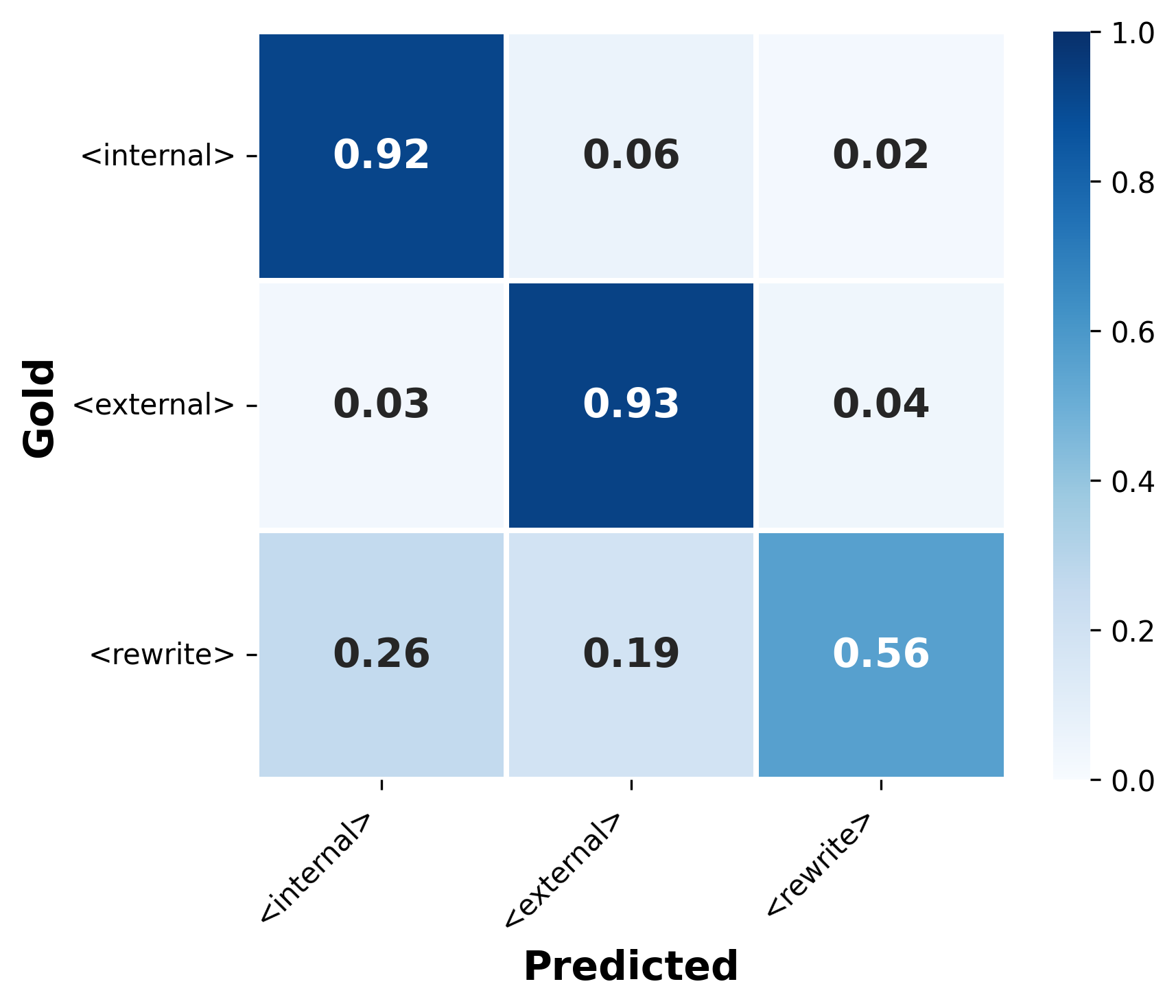}
    \caption{Bio-acoustic QA}
    \label{fig:cm-part1}
  \end{subfigure}\hfill
  \begin{subfigure}[t]{0.32\linewidth}
    \centering
    \includegraphics[width=\linewidth]{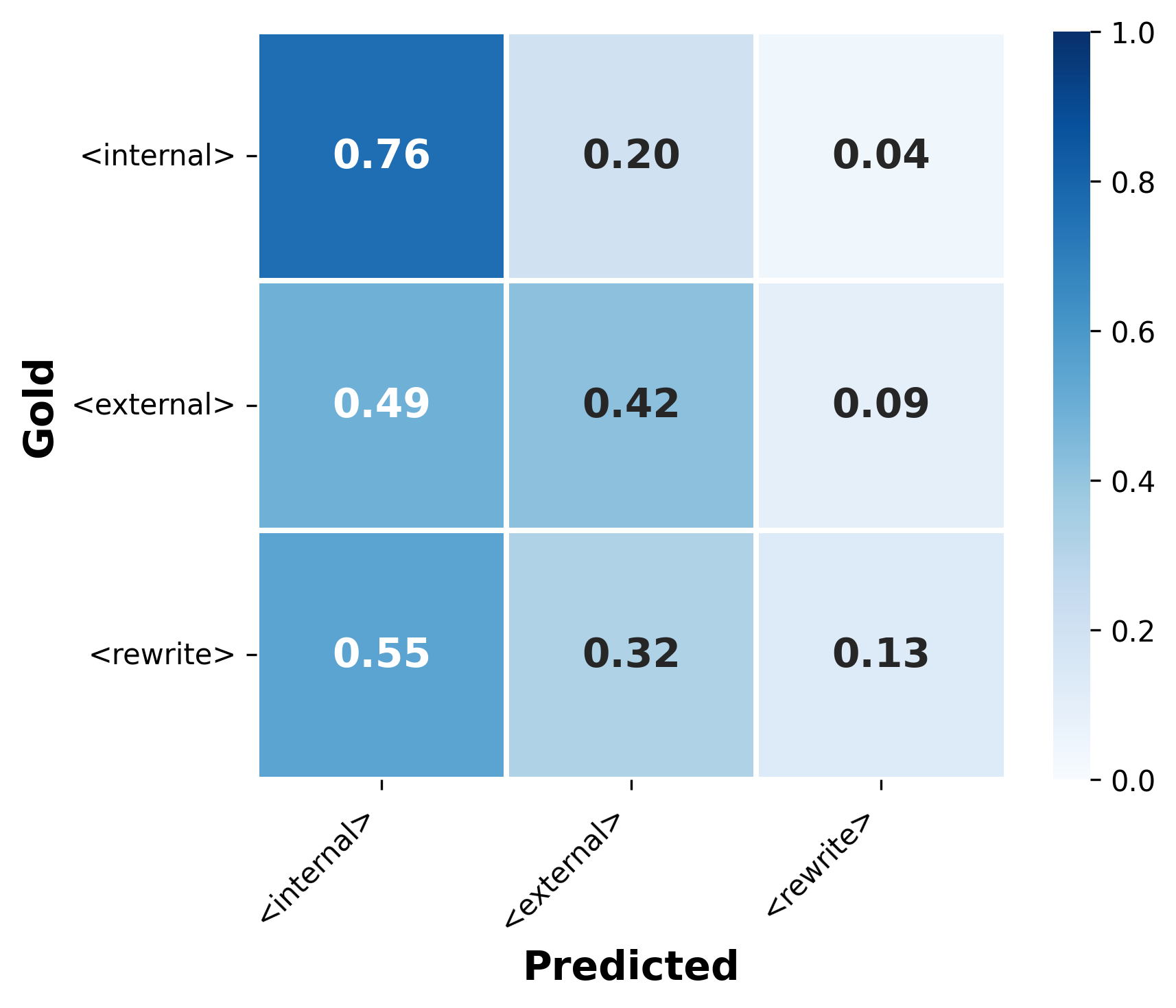}
    \caption{Soundscape QA}
    \label{fig:cm-part2}
  \end{subfigure}\hfill
  \begin{subfigure}[t]{0.32\linewidth}
    \centering
    \includegraphics[width=\linewidth]{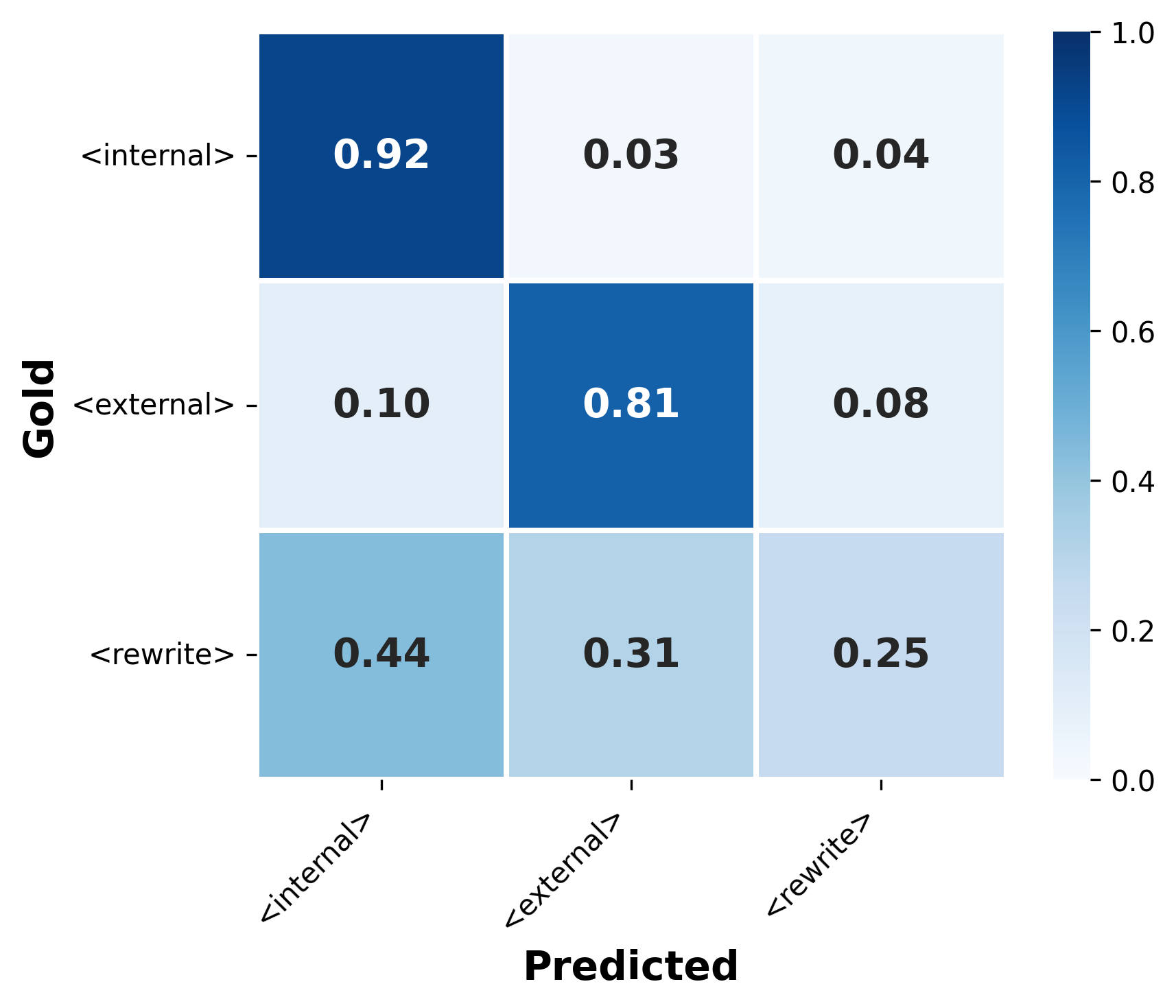}
    \caption{Complex QA}
    \label{fig:cm-part3}
  \end{subfigure}
  \caption{Confusion matrices of Speech-Hands' agentic action execution for audio QA and reasoning three subsets on (a) bio-acoustic QA, (b) temporal and sound event QA, and (c) complex audio information QA. }
  \label{fig:cm-all}
\end{figure*}

\subsection{Confusion Analysis In AudioQA}

\begin{table}[h!]
\centering
\small
\setlength{\tabcolsep}{6pt}
\begin{tabular}{lcc}
\toprule
\textbf{Subset}  & \textbf{<in>/<ex>/<re>} \\
\midrule
Bio-acoustic QA  & 106/75/43\\
Soundscape QA    & 343/185/81\\
Complex QA       & 978/555/100\\
\bottomrule
\end{tabular}
\caption{Oracle Statistics of the three DCASE2025 AudioQA test sets used in our experiments.}
\label{tab:audioqa_oracle}
\end{table}

We analyze the confusion matrix over the three subsets under the w/ multiple sampling setup (the oracle token distribution is shown in Table~\ref{tab:audioqa_oracle}).
As shown in Figure~\ref{fig:cm-all}, the confusion between \texttt{<internal>} and \texttt{<external>} remains relatively low, suggesting that the model can effectively distinguish between them, especially in the Complex QA subset. However, the Soundscape subset shows slightly increased overlap, possibly due to the difficulty of Soundscape compared with other parts, also their original performances (Qwen 56.32 v.s. Flamingo 57.31) are quite closed, leading to much smaller sets of \texttt{<external>}.
Besides, the highest confusion still lies with the \texttt{<rewrite>} class. As summarized in Table~\ref{tab:audioqa_stats}, the number of training tokens labeled as \texttt{<rewrite>} is significantly smaller across all subsets. Such sparsity likely limits the model’s tendency to generalize the rewriting behavior during generation, but when it generates \texttt{<rewrite>}, the accuracy is quite high and robust as shown in the Figure.
These findings highlight that even with imbalanced actions in training, the current F1 scores can still reliably improve the final performance, emphsizing the effectiveness of action tokens. 
\color{black}{}


\subsection{Category Analysis and Case Study}\label{cases}
To provide systematic insight beyond individual cases, during our prelimiary, we categorize three primary failure modes observed in ASR when Speech-Hands is absent. These categories further motivate our design choice of action tokens: (1) \textbf{External-induced Misguidance:} the internal omni-model is correct, but the model is misled by an error-heavy external hypothesis. This is the most damaging failure mode and directly motivates the action-token arbitration design;
(2) \textbf{Over-correction:} LLMs tend to “complete” disfluent or partially heard speech, introducing hallucinated insertions or semantic expansions;
(3) \textbf{Undercorrection in Dual-Failure:} when both internal and external predictions are wrong, the omni model often selects one erroneous hypothesis instead of generating a rewrite, due to the lack of an explicit rewrite mechanism.
These observations align with the high-precision behavior of the \texttt{<rewrite>} token and explain the improvements: the model selectively activates \texttt{<rewrite>} only when necessary.

\color{black}{}
We present illustrative examples (more in Appendix~\ref{more_cases} and~\ref{more_cases2} ) to show how the model makes selective use of internal, external, and rewrite hypotheses across modalities.
In the \textsc{AudioQA Case}, both the internal and external models predict ``B. Thunderstorm'', likely influenced by surface acoustic features such as low-frequency rumbling. The rewrite path, however, generates ``C. Forest fire'', which aligns with the ground truth, demonstrating the strong influence of \texttt{<rewrite>} that even two models have the same first-pass prediction.

In contrast, the \textsc{ASR Case} reveals a different decision dynamic. Although the rewrite produces a more fluent variant, the model opts to retain the baseline hypothesis, judging the original phrase as sufficiently accurate. This indicates the model’s ability to avoid overcorrection issue in prior GER researches when input ambiguity is low. These examples underscore the benefits of explicit action tokens: the model can either rely on internal or external model predictions or revise them when necessary, yielding both flexibility and robustness in audio tasks.

\begin{casecard}[\textcolor{blue}{AudioQA  Inference-Time Case Study}]
\textbf{Q.} Based on the audio, which natural phenomenon could be occurring?\\
A. Earthquake B. Thunderstorm \\C. Forest fire D. Snowstorm

\vspace{3pt}
\textbf{Internal pred:} B. Thunderstorm \\
\textbf{External pred:} B. Thunderstorm

\vspace{3pt}
\begin{qaanswer}
\textbf{Pred:} \texttt{<rewrite>} C. Forest fire ( \textcolor{green}{\cmark})
\end{qaanswer}
\end{casecard}

\begin{casecard}[\textcolor{blue}{ASR Inference-Time Case Study}]

\textbf{Internal:} you in the way marguerite but how \\
\textbf{External:} you ll in the way marguerite but how \\
\textbf{Rewrite:} you are in the way marguerite but how

\vspace{3pt}
\begin{qaanswer}
\textbf{Pred:} \texttt{<internal>} you in the way marguerite but how (\textcolor{green}{\cmark})
\end{qaanswer}
\end{casecard}

\section{Conclusion}

In this work, we proposed a learnable voice-agentic framework Speech-Hands for teaching omni models when to trust itself versus when to consult external audio perception. By casting the problem with explicit \texttt{<internal>}, \texttt{<external>}, and \texttt{<rewrite>} action tokens, our experimental results across AudioQA and ASR benchmarks demonstrate strong performance improvements beyond strong baselines, especially when direct finetuning and GER training fail, Speech-Hands can still robustly generate the best prediction. 

This framework also benifits the interpretability in analysis, the model achieves high F1 scores for both \texttt{<internal>} and \texttt{<external>} tokens, even under imbalanced training conditions. While the \texttt{<rewrite>} token is rarer, its precision notably exceeds recall, indicating that the model can accurately identify necessary rewrites when it does trigger them.
Overall, our method offers an effective framework to inject explicit actions into agent decision, toward reliable audio intelligence.

\section*{Limitations}
\label{sec:limitations}

Despite promising results, our study presents several limitations that offer avenues for future exploration.

\paragraph{Token imbalance and rewrite sparsity.}
Our training data exhibits an inherent imbalance across action tokens (\texttt{<internal>}, \texttt{<external>}, \texttt{<rewrite>}). While both \texttt{<internal>} and \texttt{<external>} achieve high F1 scores, \texttt{<rewrite>} remains under-trained on many datasets. This sparsity partly reflects that certain audio QA datasets rarely require rewriting but this contextual information sparsity also reveals a modeling challenge. Future work may explore more principled strategies for balancing token distribution or adaptively reshaping decision boundaries, especially under varying persona settings or task configurations.

\paragraph{Limited ASR training subset.}
Our current ASR experiments are trained on a restricted subset of data. While the model already achieves strong performance, it likely underutilizes the available signal. Scaling up training with larger ASR datasets or augmenting with synthetic audio variants may unlock further gains.

\paragraph{No exploration of transfer or multi-external setups.}
We do not yet study transfer capabilities. For example, training with one external ASR model and testing with another. Moreover, our current system only accepts a single external prediction. Extending the framework to handle multiple external models, each represented by distinct decision tokens, could significantly improve robustness and enable broader deployment across diverse real-world pipeline

\section*{Acknowledgments}
This paper is also partially supported by the Research and Development Center for Large Language Models, National Institute of Informatics (NII LLMC). We appreciate the H100 DGX and LPU support from NVIDIA's Taiwan R\&D Center.

\bibliography{custom}

@article{selman1974structural,
  title={A structural-developmental analysis of levels of role taking in middle childhood},
  author={Selman, Robert L and Byrne, Diane F},
  journal={Child development},
  pages={803--806},
  year={1974},
  publisher={JSTOR}
}

@article{sakshi2024mmau,
  title={Mmau: A massive multi-task audio understanding and reasoning benchmark},
  author={Sakshi, S and Tyagi, Utkarsh and Kumar, Sonal and Seth, Ashish and Selvakumar, Ramaneswaran and Nieto, Oriol and Duraiswami, Ramani and Ghosh, Sreyan and Manocha, Dinesh},
  journal={arXiv preprint arXiv:2410.19168},
  year={2024}
}

@inproceedings{wan2025speechiq,
  title={SpeechIQ: Speech-agentic intelligence quotient across cognitive levels in voice understanding by large language models},
  author={Wan, Zhen and Yang, Chao-Han Huck and Yu, Yahan and Tian, Jinchuan and Li, Sheng and Hu, Ke and Chen, Zhehuai and Watanabe, Shinji and Cheng, Fei and Chu, Chenhui and others},
  booktitle={Proceedings of the 63rd Annual Meeting of the Association for Computational Linguistics (Volume 1: Long Papers)},
  pages={30381--30398},
  year={2025}
}

@inproceedings{lin2025neko,
  title={Neko: Cross-modality post-recognition error correction with tasks-guided mixture-of-experts language model},
  author={Lin, Yen-Ting and Chen, Zhehuai and {\.Z}elasko, Piotr and Wan, Zhen and Yang, Xuesong and Chen, Zih-Ching and Puvvada, Krishna C and Hu, Ke and Fu, Szu-Wei and Chiu, Jun Wei and others},
  booktitle={Proceedings of the 63rd Annual Meeting of the Association for Computational Linguistics (Volume 6: Industry Track)},
  pages={222--236},
  year={2025}
}

@inproceedings{sayigh2016watkins,
  title={The Watkins marine mammal sound database: an online, freely accessible resource},
  author={Sayigh, Laela and Daher, Mary Ann and Allen, Julie and Gordon, Helen and Joyce, Katherine and Stuhlmann, Claire and Tyack, Peter},
  booktitle={Proceedings of Meetings on Acoustics},
  number={1},
  pages={040013},
  year={2016},
  organization={Acoustical Society of America}
}

@article{ma2025mmar,
  title={MMAR: A Challenging Benchmark for Deep Reasoning in Speech, Audio, Music, and Their Mix},
  author={Ma, Ziyang and Ma, Yinghao and Zhu, Yanqiao and Yang, Chen and Chao, Yi-Wen and Xu, Ruiyang and Chen, Wenxi and Chen, Yuanzhe and Chen, Zhuo and Cong, Jian and others},
  journal={arXiv preprint arXiv:2505.13032},
  year={2025}
}

@techreport{Li_mlpxc_2025,
    author = "Li, Gang and Liu, Jizhong and Dinkel, Heinrich and Niu, Yadong and Sun, Xingwei and Wang, Tianzi and Zhang, Junbo and Luan, Jian",
    title = "MIAQA Submission for DCASE 2025 Challenge Task 5: A Reinforcement Learning Driven Audio Question Answering Method",
    institution = "DCASE2025 Challenge",
    year = "2025",
    month = "June"
}

@inproceedings{songmodularized,
  title={Modularized Self-Reflected Video Reasoner for Multimodal LLM with Application to Video Question Answering},
  author={Song, Zihan and Wang, Xin and Qian, Zi and Chen, Hong and Huang, Longtao and Xue, Hui and Zhu, Wenwu},
  booktitle={Forty-second International Conference on Machine Learning},
  year={2025}
}

@article{yang2025humanomniv2,
  title={HumanOmniV2: From Understanding to Omni-Modal Reasoning with Context},
  author={Yang, Qize and Yao, Shimin and Chen, Weixuan and Fu, Shenghao and Bai, Detao and Zhao, Jiaxing and Sun, Boyuan and Yin, Bowen and Wei, Xihan and Zhou, Jingren},
  journal={arXiv preprint arXiv:2506.21277},
  year={2025}
}

@inproceedings{hu-etal-2025-investigating,
    title = "Investigating and Enhancing Vision-Audio Capability in Omnimodal Large Language Models",
    author = "Hu, Rui  and
      Qiu, Delai  and
      Wei, Shuyu  and
      Zhang, Jiaming  and
      Wang, Yining  and
      Liu, Shengping  and
      Sang, Jitao",
    editor = "Che, Wanxiang  and
      Nabende, Joyce  and
      Shutova, Ekaterina  and
      Pilehvar, Mohammad Taher",
    booktitle = "Findings of the Association for Computational Linguistics: ACL 2025",
    month = jul,
    year = "2025",
    address = "Vienna, Austria",
    publisher = "Association for Computational Linguistics",
    url = "https://aclanthology.org/2025.findings-acl.389/",
    doi = "10.18653/v1/2025.findings-acl.389",
    pages = "7452--7463",
    ISBN = "979-8-89176-256-5"
}

@inproceedings{yang2023generative,
  title={Generative speech recognition error correction with large language models and task-activating prompting},
  author={Yang, Chao-Han Huck and Gu, Yile and Liu, Yi-Chieh and Ghosh, Shalini and Bulyko, Ivan and Stolcke, Andreas},
  booktitle={2023 IEEE Automatic Speech Recognition and Understanding Workshop (ASRU)},
  pages={1--8},
  year={2023},
  organization={IEEE}
}

@misc{radford2022robustspeechrecognitionlargescale,
      title={Robust Speech Recognition via Large-Scale Weak Supervision}, 
      author={Alec Radford and Jong Wook Kim and Tao Xu and Greg Brockman and Christine McLeavey and Ilya Sutskever},
      year={2022},
      eprint={2212.04356},
      archivePrefix={arXiv},
      primaryClass={eess.AS},
      url={https://arxiv.org/abs/2212.04356}, 
}

@inbook{Hernandez_2018,
   title={TED-LIUM 3: Twice as Much Data and Corpus Repartition for Experiments on Speaker Adaptation},
   ISBN={9783319995793},
   ISSN={1611-3349},
   url={http://dx.doi.org/10.1007/978-3-319-99579-3_21},
   DOI={10.1007/978-3-319-99579-3_21},
   booktitle={Speech and Computer},
   publisher={Springer International Publishing},
   author={Hernandez, François and Nguyen, Vincent and Ghannay, Sahar and Tomashenko, Natalia and Estève, Yannick},
   year={2018},
   pages={198–208} }

@misc{li2025baichuanomni15technicalreport,
      title={Baichuan-Omni-1.5 Technical Report}, 
      author={Yadong Li and Jun Liu and Tao Zhang and Tao Zhang and Song Chen and Tianpeng Li and Zehuan Li and Lijun Liu and Lingfeng Ming and Guosheng Dong and Da Pan and Chong Li and others},
      year={2025},
      eprint={2501.15368},
      archivePrefix={arXiv},
      primaryClass={cs.CL},
      url={https://arxiv.org/abs/2501.15368}, 
}

@article{Calcus2024,
  author    = {Calcus, A.},
  title     = {Development of auditory scene analysis: a mini-review},
  journal   = {Frontiers in Human Neuroscience},
  volume    = {18},
  pages     = {1352247},
  year      = {2024},
  month     = {Mar 12},
  doi       = {10.3389/fnhum.2024.1352247},
  pmid      = {38532788},
  pmcid     = {PMC10963424}
}

@article{xu2025qwen2,
  title={Qwen2. 5-omni technical report},
  author={Xu, Jin and Guo, Zhifang and He, Jinzheng and Hu, Hangrui and He, Ting and Bai, Shuai and Chen, Keqin and Wang, Jialin and Fan, Yang and Dang, Kai and others},
  journal={arXiv preprint arXiv:2503.20215},
  year={2025}
}

@inproceedings{radford2023robust,
  title={Robust speech recognition via large-scale weak supervision},
  author={Radford, Alec and Kim, Jong Wook and Xu, Tao and Brockman, Greg and McLeavey, Christine and Sutskever, Ilya},
  booktitle={International conference on machine learning},
  pages={28492--28518},
  year={2023},
  organization={PMLR}
}

@article{Galantucci2006,
  author  = {Galantucci, B. and Fowler, C. A. and Turvey, M. T.},
  title   = {{The motor theory of speech perception reviewed}},
  journal = {Psychonomic Bulletin \& Review},
  year    = {2006},
  month   = {jun},
  volume  = {13},
  number  = {3},
  pages   = {361--377},
  doi     = {10.3758/bf03193857},
  pmid    = {17048719},
  pmcid   = {PMC2746041},
  note    = {Erratum in: Psychon Bull Rev. 2006 Aug;13(4):742}
}

@incollection{Nelson1990,
  author    = {Nelson, Thomas O.},
  title     = {{Metamemory: A Theoretical Framework and New Findings}},
  booktitle = {Psychology of Learning and Motivation},
  editor    = {Bower, Gordon H.},
  publisher = {Academic Press},
  year      = {1990},
  volume    = {26},
  pages     = {125--173},
  doi       = {10.1016/S0079-7421(08)60053-5},
  url       = {https://www.sciencedirect.com/science/article/pii/S0079742108600535},
  isbn      = {9780125433266},
  issn      = {0079-7421},
}

@article{Lebiere2011,
  author  = {Lebiere, C. and Anderson, J. R.},
  title   = {{Cognitive Constraints on Decision Making under Uncertainty}},
  journal = {Frontiers in Psychology},
  year    = {2011},
  month   = {nov},
  volume  = {2},
  pages   = {305},
  doi     = {10.3389/fpsyg.2011.00305},
  pmid    = {22110458},
  pmcid   = {PMC3216026}
}

@article{Kaiser2021,
  author  = {Kaiser, M. and Senkowski, D. and Keil, J.},
  title   = {{Mediofrontal theta-band oscillations reflect top-down influence in the ventriloquist illusion}},
  journal = {Human Brain Mapping},
  year    = {2021},
  volume  = {42},
  number  = {2},
  pages   = {452--466},
  month   = {feb},
  doi     = {10.1002/hbm.25236},
  pmid    = {33617132},
  pmcid   = {PMC7775991},
  note    = {Epub 2020 Oct 14}
}

@article{xu2025qwen3,
  title={Qwen3-Omni Technical Report},
  author={Xu, Jin and Guo, Zhifang and Hu, Hangrui and Chu, Yunfei and Wang, Xiong and He, Jinzheng and Wang, Yuxuan and Shi, Xian and He, Ting and Zhu, Xinfa and others},
  journal={arXiv preprint arXiv:2509.17765},
  year={2025}
}

@article{goel2024omcat,
  title={Omcat: Omni context aware transformer},
  author={Goel, Arushi and Sapra, Karan and Le, Matthieu and Valle, Rafael and Tao, Andrew and Catanzaro, Bryan},
  journal={arXiv preprint arXiv:2410.12109},
  year={2024}
}

@article{abouelenin2025phi,
  title={Phi-4-mini technical report: Compact yet powerful multimodal language models via mixture-of-loras},
  author={Abouelenin, Abdelrahman and Ashfaq, Atabak and Atkinson, Adam and Awadalla, Hany and Bach, Nguyen and Bao, Jianmin and Benhaim, Alon and Cai, Martin and Chaudhary, Vishrav and Chen, Congcong and others},
  journal={arXiv preprint arXiv:2503.01743},
  year={2025}
}

@article{renze2024self,
  title={Self-reflection in llm agents: Effects on problem-solving performance},
  author={Renze, Matthew and Guven, Erhan},
  journal={arXiv preprint arXiv:2405.06682},
  year={2024}
}

@article{madaan2023self,
  title={Self-refine: Iterative refinement with self-feedback},
  author={Madaan, Aman and Tandon, Niket and Gupta, Prakhar and Hallinan, Skyler and Gao, Luyu and Wiegreffe, Sarah and Alon, Uri and Dziri, Nouha and Prabhumoye, Shrimai and Yang, Yiming and others},
  journal={Advances in Neural Information Processing Systems},
  volume={36},
  pages={46534--46594},
  year={2023}
}

@article{cheng2024vision,
  title={Vision-language models can self-improve reasoning via reflection},
  author={Cheng, Kanzhi and Li, Yantao and Xu, Fangzhi and Zhang, Jianbing and Zhou, Hao and Liu, Yang},
  journal={arXiv preprint arXiv:2411.00855},
  year={2024}
}

@article{fang2025fewer,
  title={Fewer Hallucinations, More Verification: A Three-Stage LLM-Based Framework for ASR Error Correction},
  author={Fang, Yangui and Cheng, Baixu and Peng, Jing and Li, Xu and Xi, Yu and Zhang, Chengwei and Zhong, Guohui},
  journal={arXiv preprint arXiv:2505.24347},
  year={2025}
}

@misc{xu2025qwen25omnitechnicalreport,
      title={Qwen2.5-Omni Technical Report}, 
      author={Jin Xu and Zhifang Guo and Jinzheng He and Hangrui Hu and Ting He and Shuai Bai and Keqin Chen and Jialin Wang and Yang Fan and Kai Dang and Bin Zhang and Xiong Wang and Yunfei Chu and Junyang Lin},
      year={2025},
      eprint={2503.20215},
      archivePrefix={arXiv},
      primaryClass={cs.CL},
      url={https://arxiv.org/abs/2503.20215}, 
}

@misc{xie2024miniomnilanguagemodelshear,
      title={Mini-Omni: Language Models Can Hear, Talk While Thinking in Streaming}, 
      author={Zhifei Xie and Changqiao Wu},
      year={2024},
      eprint={2408.16725},
      archivePrefix={arXiv},
      primaryClass={cs.AI},
      url={https://arxiv.org/abs/2408.16725}, 
}

@misc{openai2024gpt4technicalreport,
      title={GPT-4 Technical Report}, 
      author={OpenAI and Josh Achiam and Steven Adler and Sandhini Agarwal and Lama Ahmad and Ilge Akkaya and Florencia Leoni Aleman and Diogo Almeida and Janko Altenschmidt and Sam Altman and Shyamal Anadkat and Red Avila and Igor Babuschkin and Suchir Balaji and Valerie Balcom and Paul Baltescu and Haiming Bao and Mohammad Bavarian and Jeff Belgum and Irwan Bello and Jake Berdine and Gabriel Bernadett-Shapiro and Christopher Berner and Lenny Bogdonoff and Oleg Boiko and Madelaine Boyd and Anna-Luisa Brakman and Greg Brockman and Tim Brooks and Miles Brundage and Kevin Button and Trevor Cai and Rosie Campbell and Andrew Cann and Brittany Carey and Chelsea Carlson and Rory Carmichael and Brooke Chan and Che Chang and Fotis Chantzis and Derek Chen and Sully Chen and Ruby Chen and Jason Chen and Mark Chen and Ben Chess and Chester Cho and Casey Chu and Hyung Won Chung and Dave Cummings and Jeremiah Currier and Yunxing Dai and Cory Decareaux and Thomas Degry and Noah Deutsch and Damien Deville and Arka Dhar and David Dohan and Steve Dowling and Sheila Dunning and Adrien Ecoffet and Atty Eleti and Tyna Eloundou and David Farhi and Liam Fedus and Niko Felix and Simón Posada Fishman and Juston Forte and Isabella Fulford and Leo Gao and Elie Georges and Christian Gibson and Vik Goel and Tarun Gogineni and Gabriel Goh and Rapha Gontijo-Lopes and Jonathan Gordon and Morgan Grafstein and Scott Gray and Ryan Greene and Joshua Gross and Shixiang Shane Gu and Yufei Guo and Chris Hallacy and Jesse Han and Jeff Harris and Yuchen He and Mike Heaton and Johannes Heidecke and Chris Hesse and Alan Hickey and Wade Hickey and Peter Hoeschele and Brandon Houghton and Kenny Hsu and Shengli Hu and Xin Hu and Joost Huizinga and Shantanu Jain and Shawn Jain and Joanne Jang and Angela Jiang and Roger Jiang and Haozhun Jin and Denny Jin and Shino Jomoto and Billie Jonn and Heewoo Jun and Tomer Kaftan and Łukasz Kaiser and Ali Kamali and Ingmar Kanitscheider and Nitish Shirish Keskar and Tabarak Khan and Logan Kilpatrick and Jong Wook Kim and Christina Kim and Yongjik Kim and Jan Hendrik Kirchner and Jamie Kiros and Matt Knight and Daniel Kokotajlo and Łukasz Kondraciuk and Andrew Kondrich and Aris Konstantinidis and Kyle Kosic and Gretchen Krueger and Vishal Kuo and Michael Lampe and Ikai Lan and Teddy Lee and Jan Leike and Jade Leung and Daniel Levy and Chak Ming Li and Rachel Lim and Molly Lin and Stephanie Lin and Mateusz Litwin and Theresa Lopez and Ryan Lowe and Patricia Lue and Anna Makanju and Kim Malfacini and Sam Manning and Todor Markov and Yaniv Markovski and Bianca Martin and Katie Mayer and Andrew Mayne and Bob McGrew and Scott Mayer McKinney and Christine McLeavey and Paul McMillan and Jake McNeil and David Medina and Aalok Mehta and Jacob Menick and Luke Metz and Andrey Mishchenko and Pamela Mishkin and Vinnie Monaco and Evan Morikawa and Daniel Mossing and Tong Mu and Mira Murati and Oleg Murk and David Mély and Ashvin Nair and Reiichiro Nakano and Rajeev Nayak and Arvind Neelakantan and Richard Ngo and Hyeonwoo Noh and Long Ouyang and Cullen O'Keefe and Jakub Pachocki and Alex Paino and Joe Palermo and Ashley Pantuliano and Giambattista Parascandolo and Joel Parish and Emy Parparita and Alex Passos and Mikhail Pavlov and Andrew Peng and Adam Perelman and Filipe de Avila Belbute Peres and Michael Petrov and Henrique Ponde de Oliveira Pinto and Michael and Pokorny and Michelle Pokrass and Vitchyr H. Pong and Tolly Powell and Alethea Power and Boris Power and Elizabeth Proehl and Raul Puri and Alec Radford and Jack Rae and Aditya Ramesh and Cameron Raymond and Francis Real and Kendra Rimbach and Carl Ross and Bob Rotsted and Henri Roussez and Nick Ryder and Mario Saltarelli and Ted Sanders and Shibani Santurkar and Girish Sastry and Heather Schmidt and David Schnurr and John Schulman and Daniel Selsam and Kyla Sheppard and Toki Sherbakov and Jessica Shieh and Sarah Shoker and Pranav Shyam and Szymon Sidor and Eric Sigler and Maddie Simens and Jordan Sitkin and Katarina Slama and Ian Sohl and Benjamin Sokolowsky and Yang Song and Natalie Staudacher and Felipe Petroski Such and Natalie Summers and Ilya Sutskever and Jie Tang and Nikolas Tezak and Madeleine B. Thompson and Phil Tillet and Amin Tootoonchian and Elizabeth Tseng and Preston Tuggle and Nick Turley and Jerry Tworek and Juan Felipe Cerón Uribe and Andrea Vallone and Arun Vijayvergiya and Chelsea Voss and Carroll Wainwright and Justin Jay Wang and Alvin Wang and Ben Wang and Jonathan Ward and Jason Wei and CJ Weinmann and Akila Welihinda and Peter Welinder and Jiayi Weng and Lilian Weng and Matt Wiethoff and Dave Willner and Clemens Winter and Samuel Wolrich and Hannah Wong and Lauren Workman and Sherwin Wu and Jeff Wu and Michael Wu and Kai Xiao and Tao Xu and Sarah Yoo and Kevin Yu and Qiming Yuan and Wojciech Zaremba and Rowan Zellers and Chong Zhang and Marvin Zhang and Shengjia Zhao and Tianhao Zheng and Juntang Zhuang and William Zhuk and Barret Zoph},
      year={2024},
      eprint={2303.08774},
      archivePrefix={arXiv},
      primaryClass={cs.CL},
      url={https://arxiv.org/abs/2303.08774}, 
}

@article{7280661f53654898abc5962f926ba81a,
title = "Unleashing the killer corpus: experiences in creating the multi-everything AMI Meeting Corpus",
author = "Jean Carletta",
year = "2007",
doi = "10.1007/s10579-007-9040-x",
language = "English",
volume = "41",
pages = "181--190",
journal = "Language Resources and Evaluation",
issn = "1574-020X",
publisher = "Springer",
number = "2",
}

@inproceedings{wang-etal-2021-voxpopuli,
    title = "{V}ox{P}opuli: A Large-Scale Multilingual Speech Corpus for Representation Learning, Semi-Supervised Learning and Interpretation",
    author = "Wang, Changhan  and
      Riviere, Morgane  and
      Lee, Ann  and
      Wu, Anne  and
      Talnikar, Chaitanya  and
      Haziza, Daniel  and
      Williamson, Mary  and
      Pino, Juan  and
      Dupoux, Emmanuel",
    editor = "Zong, Chengqing  and
      Xia, Fei  and
      Li, Wenjie  and
      Navigli, Roberto",
    booktitle = "Proceedings of the 59th Annual Meeting of the Association for Computational Linguistics and the 11th International Joint Conference on Natural Language Processing (Volume 1: Long Papers)",
    month = aug,
    year = "2021",
    address = "Online",
    publisher = "Association for Computational Linguistics",
    url = "https://aclanthology.org/2021.acl-long.80/",
    doi = "10.18653/v1/2021.acl-long.80",
    pages = "993--1003"
}

@inproceedings{Chen_2021, series={interspeech\_2021},
   title={GigaSpeech: An Evolving, Multi-Domain ASR Corpus with 10,000 Hours of Transcribed Audio},
   url={http://dx.doi.org/10.21437/Interspeech.2021-1965},
   DOI={10.21437/interspeech.2021-1965},
   booktitle={Interspeech 2021},
   publisher={ISCA},
   author={Chen, Guoguo and Chai, Shuzhou and Wang, Guan-Bo and Du, Jiayu and Zhang, Wei-Qiang and Weng, Chao and Su, Dan and Povey, Daniel and Trmal, Jan and Zhang, Junbo and Jin, Mingjie and Khudanpur, Sanjeev and Watanabe, Shinji and Zhao, Shuaijiang and Zou, Wei and Li, Xiangang and Yao, Xuchen and Wang, Yongqing and You, Zhao and Yan, Zhiyong},
   year={2021},
   month=aug, collection={interspeech_2021} }

@inproceedings{oneill21_interspeech,
  title     = {SPGISpeech: 5,000 Hours of Transcribed Financial Audio for Fully Formatted End-to-End Speech Recognition},
  author    = {Patrick K. O’Neill and Vitaly Lavrukhin and Somshubra Majumdar and Vahid Noroozi and Yuekai Zhang and Oleksii Kuchaiev and Jagadeesh Balam and Yuliya Dovzhenko and Keenan Freyberg and Michael D. Shulman and Boris Ginsburg and Shinji Watanabe and Georg Kucsko},
  year      = {2021},
  booktitle = {Interspeech 2021},
  pages     = {1434--1438},
  doi       = {10.21437/Interspeech.2021-1860},
  issn      = {2958-1796},
}

@INPROCEEDINGS{7178964,
  author={Panayotov, Vassil and Chen, Guoguo and Povey, Daniel and Khudanpur, Sanjeev},
  booktitle={2015 IEEE International Conference on Acoustics, Speech and Signal Processing (ICASSP)}, 
  title={Librispeech: An ASR corpus based on public domain audio books}, 
  year={2015},
  volume={},
  number={},
  pages={5206-5210},
  doi={10.1109/ICASSP.2015.7178964}}

@misc{yang2025multidomainaudioquestionanswering,
      title={Multi-Domain Audio Question Answering Toward Acoustic Content Reasoning in The DCASE 2025 Challenge}, 
      author={Chao-Han Huck Yang and Sreyan Ghosh and Qing Wang and Jaeyeon Kim and Hengyi Hong and Sonal Kumar and Guirui Zhong and Zhifeng Kong and S Sakshi and Vaibhavi Lokegaonkar and Oriol Nieto and Ramani Duraiswami and Dinesh Manocha and Gunhee Kim and Jun Du and Rafael Valle and Bryan Catanzaro},
      year={2025},
      eprint={2505.07365},
      archivePrefix={arXiv},
      primaryClass={cs.SD},
      url={https://arxiv.org/abs/2505.07365}, 
}

@misc{sekoyan2025canary1bv2parakeettdt06bv3efficient,
      title={Canary-1B-v2 \& Parakeet-TDT-0.6B-v3: Efficient and High-Performance Models for Multilingual ASR and AST}, 
      author={Monica Sekoyan and Nithin Rao Koluguri and Nune Tadevosyan and Piotr Zelasko and Travis Bartley and Nikolay Karpov and Jagadeesh Balam and Boris Ginsburg},
      year={2025},
      eprint={2509.14128},
      archivePrefix={arXiv},
      primaryClass={cs.CL},
      url={https://arxiv.org/abs/2509.14128}, 
}

@misc{goel2025audioflamingo3advancing,
      title={Audio Flamingo 3: Advancing Audio Intelligence with Fully Open Large Audio Language Models}, 
      author={Arushi Goel and Sreyan Ghosh and Jaehyeon Kim and Sonal Kumar and Zhifeng Kong and Sang-gil Lee and Chao-Han Huck Yang and Ramani Duraiswami and Dinesh Manocha and Rafael Valle and Bryan Catanzaro},
      year={2025},
      eprint={2507.08128},
      archivePrefix={arXiv},
      primaryClass={cs.SD},
      url={https://arxiv.org/abs/2507.08128}, 
}

@misc{microsoft2025phi4minitechnicalreportcompact,
      title={Phi-4-Mini Technical Report: Compact yet Powerful Multimodal Language Models via Mixture-of-LoRAs}, 
      author={Microsoft and : and Abdelrahman Abouelenin and Atabak Ashfaq and Adam Atkinson and Hany Awadalla and Nguyen Bach and Jianmin Bao and Alon Benhaim and Martin Cai and Vishrav Chaudhary and Congcong Chen and Dong Chen and Dongdong Chen and Junkun Chen and Weizhu Chen and Yen-Chun Chen and Yi-ling Chen and Qi Dai and Xiyang Dai and Ruchao Fan and Mei Gao and Min Gao and Amit Garg and Abhishek Goswami and Junheng Hao and Amr Hendy and Yuxuan Hu and Xin Jin and Mahmoud Khademi and Dongwoo Kim and Young Jin Kim and Gina Lee and Jinyu Li and Yunsheng Li and Chen Liang and Xihui Lin and Zeqi Lin and Mengchen Liu and Yang Liu and Gilsinia Lopez and Chong Luo and Piyush Madan and Vadim Mazalov and Arindam Mitra and Ali Mousavi and Anh Nguyen and Jing Pan and Daniel Perez-Becker and Jacob Platin and Thomas Portet and Kai Qiu and Bo Ren and Liliang Ren and Sambuddha Roy and Ning Shang and Yelong Shen and Saksham Singhal and Subhojit Som and Xia Song and Tetyana Sych and Praneetha Vaddamanu and Shuohang Wang and Yiming Wang and Zhenghao Wang and Haibin Wu and Haoran Xu and Weijian Xu and Yifan Yang and Ziyi Yang and Donghan Yu and Ishmam Zabir and Jianwen Zhang and Li Lyna Zhang and Yunan Zhang and Xiren Zhou},
      year={2025},
      eprint={2503.01743},
      archivePrefix={arXiv},
      primaryClass={cs.CL},
      url={https://arxiv.org/abs/2503.01743}, 
}

@misc{koizumi2020audiocaptioningusingpretrained,
      title={Audio Captioning using Pre-Trained Large-Scale Language Model Guided by Audio-based Similar Caption Retrieval}, 
      author={Yuma Koizumi and Yasunori Ohishi and Daisuke Niizumi and Daiki Takeuchi and Masahiro Yasuda},
      year={2020},
      eprint={2012.07331},
      archivePrefix={arXiv},
      primaryClass={eess.AS},
      url={https://arxiv.org/abs/2012.07331}, 
}

@inproceedings{zhao-etal-2023-generating,
    title = "Generating Synthetic Speech from {S}poken{V}ocab for Speech Translation",
    author = "Zhao, Jinming  and
      Haffari, Gholamreza  and
      Shareghi, Ehsan",
    editor = "Vlachos, Andreas  and
      Augenstein, Isabelle",
    booktitle = "Findings of the Association for Computational Linguistics: EACL 2023",
    month = may,
    year = "2023",
    address = "Dubrovnik, Croatia",
    publisher = "Association for Computational Linguistics",
    url = "https://aclanthology.org/2023.findings-eacl.147/",
    doi = "10.18653/v1/2023.findings-eacl.147",
    pages = "1975--1981"
}

@misc{huang2023makeanaudiotexttoaudiogenerationpromptenhanced,
      title={Make-An-Audio: Text-To-Audio Generation with Prompt-Enhanced Diffusion Models}, 
      author={Rongjie Huang and Jiawei Huang and Dongchao Yang and Yi Ren and Luping Liu and Mingze Li and Zhenhui Ye and Jinglin Liu and Xiang Yin and Zhou Zhao},
      year={2023},
      eprint={2301.12661},
      archivePrefix={arXiv},
      primaryClass={cs.SD},
      url={https://arxiv.org/abs/2301.12661}, 
}

@inproceedings{gonzales-rudzicz-2024-retrieval,
    title = "A Retrieval Augmented Approach for Text-to-Music Generation",
    author = "Gonzales, Robie  and
      Rudzicz, Frank",
    editor = "Kruspe, Anna  and
      Oramas, Sergio  and
      Epure, Elena V.  and
      Sordo, Mohamed  and
      Weck, Benno  and
      Doh, SeungHeon  and
      Won, Minz  and
      Manco, Ilaria  and
      Meseguer-Brocal, Gabriel",
    booktitle = "Proceedings of the 3rd Workshop on NLP for Music and Audio (NLP4MusA)",
    month = nov,
    year = "2024",
    address = "Oakland, USA",
    publisher = "Association for Computational Lingustics",
    url = "https://aclanthology.org/2024.nlp4musa-1.6/",
    pages = "31--36"
}

@misc{yang2023mmreactpromptingchatgptmultimodal,
      title={MM-REACT: Prompting ChatGPT for Multimodal Reasoning and Action}, 
      author={Zhengyuan Yang and Linjie Li and Jianfeng Wang and Kevin Lin and Ehsan Azarnasab and Faisal Ahmed and Zicheng Liu and Ce Liu and Michael Zeng and Lijuan Wang},
      year={2023},
      eprint={2303.11381},
      archivePrefix={arXiv},
      primaryClass={cs.CV},
      url={https://arxiv.org/abs/2303.11381}, 
}

@misc{zhang2024omagentmultimodalagentframework,
      title={OmAgent: A Multi-modal Agent Framework for Complex Video Understanding with Task Divide-and-Conquer}, 
      author={Lu Zhang and Tiancheng Zhao and Heting Ying and Yibo Ma and Kyusong Lee},
      year={2024},
      eprint={2406.16620},
      archivePrefix={arXiv},
      primaryClass={cs.CV},
      url={https://arxiv.org/abs/2406.16620}, 
}

@misc{wang2025audioagentleveragingllmsaudio,
      title={Audio-Agent: Leveraging LLMs For Audio Generation, Editing and Composition}, 
      author={Zixuan Wang and Chi-Keung Tang and Yu-Wing Tai},
      year={2025},
      eprint={2410.03335},
      archivePrefix={arXiv},
      primaryClass={cs.SD},
      url={https://arxiv.org/abs/2410.03335}, 
}

@misc{chen2025wavragaudiointegratedretrievalaugmented,
      title={WavRAG: Audio-Integrated Retrieval Augmented Generation for Spoken Dialogue Models}, 
      author={Yifu Chen and Shengpeng Ji and Haoxiao Wang and Ziqing Wang and Siyu Chen and Jinzheng He and Jin Xu and Zhou Zhao},
      year={2025},
      eprint={2502.14727},
      archivePrefix={arXiv},
      primaryClass={cs.SD},
      url={https://arxiv.org/abs/2502.14727}, 
}

@misc{sakshi2024mmaumassivemultitaskaudio,
      title={MMAU: A Massive Multi-Task Audio Understanding and Reasoning Benchmark}, 
      author={S Sakshi and Utkarsh Tyagi and Sonal Kumar and Ashish Seth and Ramaneswaran Selvakumar and Oriol Nieto and Ramani Duraiswami and Sreyan Ghosh and Dinesh Manocha},
      year={2024},
      eprint={2410.19168},
      archivePrefix={arXiv},
      primaryClass={eess.AS},
      url={https://arxiv.org/abs/2410.19168}, 
}

@misc{marinoni2024overviewl3das23challengeaudiovisual,
      title={Overview of the L3DAS23 Challenge on Audio-Visual Extended Reality}, 
      author={Christian Marinoni and Riccardo Fosco Gramaccioni and Changan Chen and Aurelio Uncini and Danilo Comminiello},
      year={2024},
      eprint={2402.09245},
      archivePrefix={arXiv},
      primaryClass={eess.AS},
      url={https://arxiv.org/abs/2402.09245}, 
}

@misc{adavanne2019multiroomreverberantdatasetsound,
      title={A multi-room reverberant dataset for sound event localization and detection}, 
      author={Sharath Adavanne and Archontis Politis and Tuomas Virtanen},
      year={2019},
      eprint={1905.08546},
      archivePrefix={arXiv},
      primaryClass={cs.SD},
      url={https://arxiv.org/abs/1905.08546}, 
}

\clearpage
\appendix

\appendix
\section{Failure Case of Multimodal GER}

\label{case_ger_fail}
\subsection{Amplifying Hallucination}
This is often a situation where external audio perception is not strong enough or N-best decoding introduces noise for GER. Especially when the omni model is encouraged to take all hypothesses in consideration.
\begin{casecard}[ASR Failure Case]

\textbf{Internal:} insane \\
\textbf{External (Whisper 5-best):} [' Gimseeinnnnnn', ' You can say.', ' Insta.', ' Wednesday.', " I'm from Phelps County, I'm gonna see what this guy's doing."] \\
\vspace{3pt}
\begin{qaanswer}
\textbf{GER Pred:}  I'm from Phelps County, I'm gonna see what this guy's doing \\
\textbf{Gold:} insane
\end{qaanswer}
\end{casecard}

\subsection{Overcorrection}
Due the language modeling nature, LLMs tend to revise the transcription to be more likea complete sentence, which sometimes caused ``overcorrection.''

\begin{casecard}[ASR Failure Case]

\textbf{Internal:} you in the way marguerite but how \\
\textbf{External (Whisper 5-best):} [' you ll in the way marguerite but how', 'you in the way marguerite but how', 'you in the way marguerite but how.', ' you in the way marguerite but how.', " you in the way marguerite but how."] \\
\vspace{3pt}
\begin{qaanswer}
\textbf{GER Pred:}  you are in the way marguerite but how \\
\textbf{Gold:} you in the way marguerite but how
\end{qaanswer}
\end{casecard}

\subsection{Prompt Ablations in Preliminary SFT}\label{app:prompt_ablation}
We conduct these experiments during early-stage exploration: for inputs we investigated both (a) audio + external whisper 5-best and (b) audio + internal 1-best + external whisper 5-best, together with four prompting strategies instructing the model to emphasize internal hypotheses, external hypotheses, audio grounding, or a balanced fusion. The average WER results on OpenASR are shown in Table~\ref{result:prompt_ablation}.

\begin{figure}[t!]
    \centering
    \includegraphics[width=\linewidth]{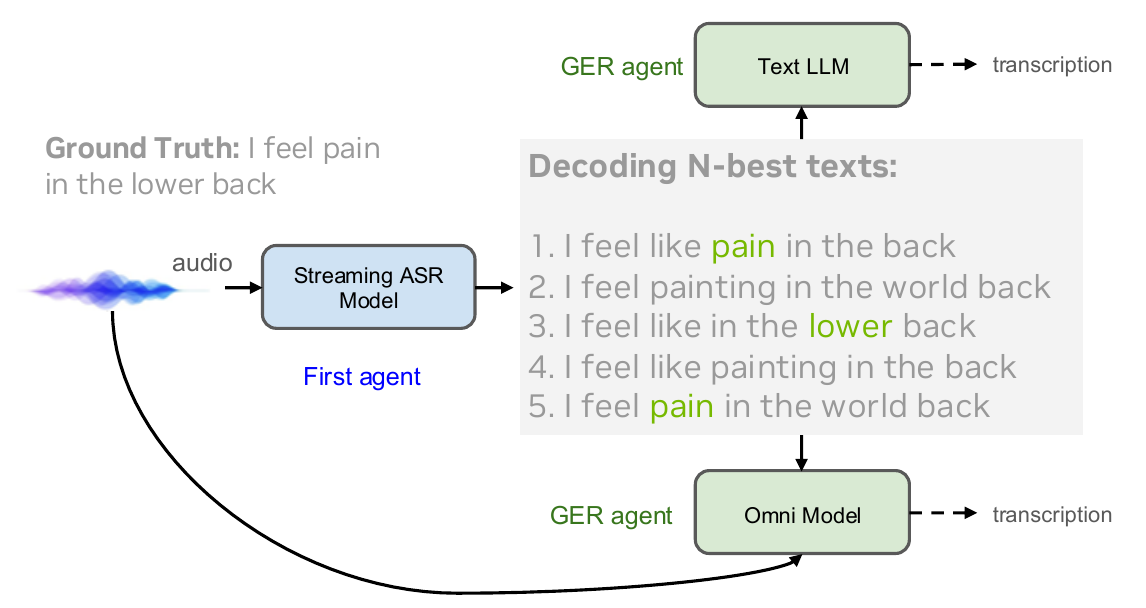}
    \caption{Text-base GER uses only ASR hypotheses. This setup fails to correct deletion or hallucination if all hypotheses are wrong. Multimodal GER include the original audio as grounding to improve error correction.}
    \label{fig:extension}
\end{figure}

\subsection{Zero-shot Qwen in Preliminary}
~\label{app:zeroshot_analysis}
We test three different zero-shot prompting strategies as shown in Table~\ref{tab:zeroshot_matrix}. We find that zero-shot decisions are highly prompt-sensitive: the model often collapses into trivial heuristics, while the balanced prompt produces unstable and inconsistent behavior across samples. The corresponding 2×2 confusion matrices show large off-diagonal mass for all zero-shot prompts, confirming that zero-shot Qwen does not perform genuine self-reflection.

\color{black}{}
\section{Prompt Template}
\label{audioqa_prompt}

Here is the prompt for AudioQA, we explicitly prompt the model to ourput \texttt{<external>} only when the internal prediction is wrong while external perception is correct.
\begin{quote}
\small
\ttfamily
You are an audio understanding model with access to three inputs:\\
(1) The original audio;\\
(2) One answer candidate generated by another model (external);\\
(3) Your own prediction (internal).\\

Your task is to decide which of the following strategies to apply:\\
- If your internal prediction is correct and acceptable, output <internal> and repeat your answer.\\
- If the external candidate is correct while your internal prediction is incorrect, output <external> and use the external answer.\\
- If all given answers are incorrect, output <rewrite> and re-answer the question correctly based only on the original audio.\\

Return the selected token (<internal>/<external>/<rewrite>) followed by your final answer.
\end{quote}

\section{Additional Agentic Studies on ASR and AudioQA Tool Calling}

\subsection{Active Perception via DSP Tool Calling}
\label{app:active_perception}

While the main {\it Speech-Hands} framework arbitrates between internal and external semantic hypotheses, our error analysis (Section~\ref{cases}) reveals that ``Dual-Failure'' modes often stem from intrinsic audio degradation. For example, background noise in Soundscape QA and poor microphone quality in AMI meetings are issues that could be mitigated through additional post-processing.

To address this, we conduct an extra mini-experiment expanding the agent's action space from purely generative decisions to \textbf{Active Perception}. We integrate open license Digital Signal Processing (DSP) tools as executable actions, allowing the model to ``clean its ears'' before transcription.

\subsection{Tool-Based Action Space}
We extend the action token set $\mathcal{A} = \{\texttt{<internal>}, \texttt{<external>}, \texttt{<rewrite>}\}$ with two signal enhancement tools:

\paragraph{Background Noise Removal (BNR) (\texttt{<tool:bnr>}).} 
This action triggers the open \textbf{NVIDIA BNR inference microservices}, designed to suppress non-speech broad-spectrum noise (e.g., traffic, sirens) while preserving emotive vocal tones. 
\begin{itemize}
    \item \textbf{Trigger Logic:} The agent emits this token when the internal audio embedding suggests high background entropy or when the initial decode contains disfluency markers like \texttt{[noise]} or \texttt{[unintelligible]}.
\end{itemize}

\paragraph{Studio Voice Restoration (\texttt{<tool:studio>}).}
This action triggers \textbf{NVIDIA Studio Voice} (via Maxine SDK), which uses generative upsampling to reconstruct high-frequency harmonics (48kHz upsampling) from low-bandwidth input.
\begin{itemize}
    \item \textbf{Trigger Logic:} Activated when the agent detects low sample rates ($<16$kHz), narrow-band telephony audio, or significant room reverberation (common in the AMI corpus).
\end{itemize}

\subsection{Experimental Results}
We evaluate this \textit{Active Speech-Hands} setup on the two most challenging subsets identified in our main results: \textbf{AMI} (noisy meetings) and \textbf{Soundscape QA} (environmental noise). As shown in Table~\ref{tab:dsp_tools}, enabling DSP tools significantly benefits these noise-sensitive tasks.

\begin{table}[h]
\centering
\small
\renewcommand{\arraystretch}{1.2}
\setlength{\tabcolsep}{5pt}
\resizebox{1.0\linewidth}{!}{
\begin{tabular}{lrr}
\toprule
\textbf{Metric / Dataset} & \textbf{Standard} & \textbf{w/ Active Tools} \\
& \textit{(Passive)} & \textit{(BNR + Studio)} \\
\midrule
\rowcolor{blue!10} \multicolumn{3}{l}{\textbf{Speech Recognition (WER $\downarrow$)}} \\
AMI (Meeting) & 15.03 & \textbf{13.41} \textcolor{darkgreen}{(-1.62)} \\
VoxPopuli (Reverb) & 6.49 & \textbf{5.82} \textcolor{darkgreen}{(-0.67)} \\
\midrule
\rowcolor{orange!10} \multicolumn{3}{l}{\textbf{Audio Reasoning (Acc \% $\uparrow$)}} \\
Soundscape QA & 59.40 & \textbf{63.15} \textcolor{darkgreen}{(+3.75)} \\
Complex QA & 85.70 & \textbf{86.44} \textcolor{darkgreen}{(+0.74)} \\
\bottomrule
\end{tabular}
}
\caption{Mini-experiment results on Active Perception. Integrating \texttt{<tool:bnr>} and \texttt{<tool:studio>} improves performance on noisy datasets (AMI) and acoustic reasoning tasks compared to the passive Speech-Hands baseline.}
\label{tab:dsp_tools}
\end{table}

\paragraph{Analysis.} 
On the AMI dataset, \texttt{<tool:studio>} successfully reconstructs fricatives lost in distant-microphone recordings, reducing WER by 1.62\%. In Soundscape QA, \texttt{<tool:bnr>} helps segregate foreground events from background clutter. 

\begin{casecard}[Active Tool Case Study (Soundscape)]
\textbf{Audio:} A faint voice speaking over loud construction drilling noise. \\
\textbf{Q.} What is the speaker asking for? \\

\vspace{3pt}
\textbf{Standard (Passive):} 
\textit{Internal:} [Noise] \\
\textit{External:} Help me with the grill. \\
\textbf{Prediction:} \texttt{<external>} Help me with the grill. (\ding{55}) \\

\vspace{3pt}
\textbf{Active (Ours):} 
\textit{Action:} \texttt{<tool:bnr>} $\rightarrow$ \textit{Cleaned Audio} \\
\textit{New Internal:} Help me with the drill. \\
\textbf{Prediction:} \texttt{<internal>} Help me with the drill. (\textcolor{green}{\cmark})
\end{casecard}

This demonstrates that for highly degraded inputs, agentic reasoning should precede perception, which decides \textit{how} to listen is as important as deciding \textit{what} was heard.

\begin{table*}[h]
\centering
\renewcommand{\arraystretch}{1.3}
\setlength{\tabcolsep}{5pt}
\resizebox{1.0\linewidth}{!}{
\begin{tabular}{lrrrrrrrr}
\toprule
\textbf{Dataset} & AMI & Tedlium & gigaspeech & spgispeech & Voxpopuli & Libri & Libri-clean & Libri-other \\
\midrule
\rowcolor{green!10}
\textbf{Sampling \#} & \multicolumn{8}{c}{Subset in Speech-Hands Training} \\
\midrule
\textbf{Train} & 20,000 & 20,000 & 9,389 & 20,000 & 20,000 & 20,000 & 20,000 & 20,000 \\
\bottomrule
\end{tabular}
}
\caption{Number of training samples used from each dataset. For GigaSpeech, only 9,389 valid samples met our filtering criteria.}
\label{subset}
\end{table*}

\section{Audio Dataset Details}
\label{data_detail}
\subsection{Dataset Details for Speech Recognition}
To ensure a fair and balanced evaluation across diverse speech corpora, we uniformly sampled up to 20,000 audio-question pairs from each dataset for training as in Table~\ref{subset}. All datasets were aligned to a consistent prompt-question-answer format to support unified multi-dataset training.
\subsection{Dataset Details for DCASE2025 AudioQA}
\begin{table}[ht!]
\centering
\small
\setlength{\tabcolsep}{6pt}
\begin{tabular}{lrrr}
\toprule
\textbf{Subset} & \textbf{\#Train / \#Dev} & \textbf{<in>/<ex>/<re>} \\
\midrule
Bio-acoustic QA & 0.7K / 0.2K & 338/234/168\\
Soundscape QA   & 1.0K / 0.6K & 604/182/252\\
Complex QA      & 6.4K / 1.6K & 4,267/1,785/391\\
\bottomrule
\end{tabular}
\caption{Statistics of the three DCASE2025 AudioQA subsets used in our experiments. \textbf{<in>/<ex>/<re>} shows the token distribution in training w/ majority sampling.}
\label{tab:audioqa_stats}
\end{table}

The DCASE2025 AudioQA benchmark comprises three complementary multiple-choice question-answering subsets, each designed to evaluate a different aspect of audio reasoning.

\subsection{Bioacoustics QA}

This subset targets perceptual and cognitive grounding in marine bioacoustics. It includes questions about 31 marine mammal species with diverse acoustic ranges, habitats, and vocalization durations. Tasks include species classification, vocalization type recognition, factual retrieval, interpretation of acoustic features, and comparative reasoning. The dataset includes approximately 0.7K training and 0.2K development QA pairs. Audio clips range in sample rate from 600 Hz to 160 kHz and in duration from 0.4 s to 625 s, allowing evaluation under highly varied acoustic conditions. All audio is sourced from the Watkins Marine Mammal Sound Database (Woods Hole Oceanographic Institution; New Bedford Whaling Museum), and usage of audio beyond the provided splits is strictly prohibited.

\subsection{Temporal Soundscapes QA}

This subset focuses on temporal reasoning over sound events, encompassing 26 event classes. Questions require identifying active sound classes, temporal ordering, timestamp estimation (onset, offset, duration), and event comparison. The subset comprises approximately 1.0K training and 0.6K development QA pairs. Audio clips are mono-channel, 10 seconds long, and sampled at 32–48 kHz. Most clips correspond to a single QA item, while a small portion supports multiple questions. All annotations are manually verified, which include event types, timestamps, and answers. Audio is sourced from NIGENS, L3DAS23 Challenge~\cite{marinoni2024overviewl3das23challengeaudiovisual}, and TAU Spatial Sound Events 2019 datasets~\cite{adavanne2019multiroomreverberantdatasetsound}.

\subsection{Complex QA (MMAU)}

This subset evaluates high-level reasoning over natural sound scenes. Each instance consists of a 10-second, 16 kHz audio clip paired with a question requiring reasoning over acoustic, temporal, and contextual cues (\textit{i.e.}, overlapping events, implied sequences, or abstract relationships.) Tasks are inspired by the MMAU Sound benchmark~\cite{sakshi2024mmaumassivemultitaskaudio} and extend its principles to more diverse soundscapes. The data include approximately 6.4K training and 1.6K development QA pairs. Audio clips are sourced from AudioSet and Mira.

\section{Cases On ASR}

We provide several examples to show the effectiveness of our Speech-Hands in ASR tasks.
\label{more_cases}
\begin{casecard}[ASR Case 1]
\textbf{Q.} What is the correct transcription of this audio? \\

\vspace{3pt}
\textbf{Internal:} water and fresh stalls to take on there \\
\textbf{External:} we've water and fresh stores to tick on there\\
\textbf{Rewrite:} we've water and fresh stores to take on there \\

\vspace{3pt}
\begin{qaanswer}
\textbf{Pred:} \texttt{<rewrite>} we've water and fresh stores to take on there \\
\textbf{Gold:} we've water and fresh stores to take on there
\end{qaanswer}
\end{casecard}

\begin{casecard}[ASR Case 2]
\textbf{Q.} What is the correct transcription of this audio? \\

\vspace{3pt}
\textbf{Internal:} then has nothing happened with a voice like a whip mr wicker said start at once \\
\textbf{External:} then as nothing happened with a voice like a whip mister wigger said start at once \\
\textbf{Rewrite:} then as nothing happened with a voice like a whip mister wuthers said start at once \\

\vspace{3pt}
\begin{qaanswer}
\textbf{Pred:} \texttt{<external>} then as nothing happened with a voice like a whip mister wigger said start at once \\
\textbf{Gold:} then as nothing happened with a voice like a whip mister wicker said start at once
\end{qaanswer}
\end{casecard}

\section{Cases On AudioQA}
\label{more_cases2}
For each token we pick up 5 examples to illustrate how action tokens influence the final correctness. For \texttt{<rewrite>} token, it works not only when the internal prediction is conflicted with the external prediction, but also show robust and accurate decision when hypothesses generate the same but wrong prediction. This is extremely important when facing challenging tasks, the omni-agent should be confident to refuse wrong hypothesses. Also for \texttt{<internal>} and \texttt{<external>} tokens that have more frequencies, they should be accurately distinguished when conflicts, and action tokens complete a good job on recognizing the true answer.
\begin{casecard}[AudioQA Case 1]
\textbf{Q.} Based on the audio, which natural phenomenon could be occurring?\\
\textit{Options:}\\
A. Earthquake \\
B. Thunderstorm \\
C. Forest fire \\
D. Snowstorm

\vspace{3pt}
\textbf{Internal pred:} B. Thunderstorm \\
\textbf{External pred:} B. Thunderstorm

\vspace{3pt}
\begin{qaanswer}
\textbf{Pred:} \texttt{<rewrite>} C. Forest fire \\
\textbf{Gold:} \texttt{<rewrite>} C. Forest fire
\end{qaanswer}
\end{casecard}

\begin{casecard}[AudioQA Case 2]
\textbf{Q.} What activity is likely occurring based on the described sounds?\\
\textit{Options:}\\
A. Driving a car \\
B. Planting a garden \\
C. Painting a picture \\
D. Cooking a meal

\vspace{3pt}
\textbf{Internal pred:} A. Driving a car \\
\textbf{External pred:} B. Planting a garden

\vspace{3pt}
\begin{qaanswer}
\textbf{Pred:} \texttt{<rewrite>} D. Cooking a meal \\
\textbf{Gold:} \texttt{<rewrite>} D. Cooking a meal
\end{qaanswer}
\end{casecard}

\begin{casecard}[AudioQA Case 3]
\textbf{Q.} Based on the audio, what is the most likely environment in which the sounds are occurring?\\
\textit{Options:}\\
A. An indoor setting \\
B. A rural countryside \\
C. A busy city street \\
D. A quiet residential neighborhood

\vspace{3pt}
\textbf{Internal pred:} A. An indoor setting \\
\textbf{External pred:} B. A rural countryside

\vspace{3pt}
\begin{qaanswer}
\textbf{Pred:} \texttt{<rewrite>} C. A busy city street \\
\textbf{Gold:} \texttt{<rewrite>} C. A busy city street
\end{qaanswer}
\end{casecard}

\begin{casecard}[AudioQA Case 4]
\textbf{Q.} What background sound is present in the audio clip?\\
\textit{Options:}\\
A. Waves crashing \\
B. Birds chirping \\
C. Bells ringing \\
D. Running car engine

\vspace{3pt}
\textbf{Internal pred:} A. Waves crashing \\
\textbf{External pred:} D. Running car engine

\vspace{3pt}
\begin{qaanswer}
\textbf{Pred:} \texttt{<rewrite>} C. Bells ringing \\
\textbf{Gold:} \texttt{<rewrite>} C. Bells ringing
\end{qaanswer}
\end{casecard}

\begin{casecard}[AudioQA Case 5]
\textbf{Q.} What might the purpose of tapping the metal object in the background be?\\
\textit{Options:}\\
A. To emphasize the speaker's instructions \\
B. To distract from the speaker's sad tone \\
C. To demonstrate its material quality \\
D. To create a rhythmic background

\vspace{3pt}
\textbf{Internal pred:} A. To emphasize the speaker's instructions \\
\textbf{External pred:} A. To emphasize the speaker's instructions

\vspace{3pt}
\begin{qaanswer}
\textbf{Pred:} \texttt{<rewrite>} C. To demonstrate its material quality \\
\textbf{Gold:} \texttt{<rewrite>} C. To demonstrate its material quality
\end{qaanswer}
\end{casecard}

\begin{casecard}[AudioQA Case 6]
\textbf{Q.} What type of mood is conveyed through the musical elements in this audio?\\
\textit{Options:}\\
A. Calm and soothing \\
B. Angry and aggressive \\
C. Joyful and uplifting \\
D. Sad and reflective

\vspace{3pt}
\textbf{Internal pred:} B. Angry and aggressive \\
\textbf{External pred:} D. Sad and reflective

\vspace{3pt}
\begin{qaanswer}
\textbf{Pred:} \texttt{<external>} D. Sad and reflective \\
\textbf{Gold:} \texttt{<external>} D. Sad and reflective
\end{qaanswer}
\end{casecard}

\begin{casecard}[AudioQA Case 7]
\textbf{Q.} Based on the audio description, what is likely happening in the background?\\
\textit{Options:}\\
A. A calm evening \\
B. A windy day \\
C. An earthquake \\
D. A quiet morning

\vspace{3pt}
\textbf{Internal pred:} B. A windy day \\
\textbf{External pred:} C. An earthquake

\vspace{3pt}
\begin{qaanswer}
\textbf{Pred:} \texttt{<external>} C. An earthquake \\
\textbf{Gold:} \texttt{<external>} C. An earthquake
\end{qaanswer}
\end{casecard}

\begin{casecard}[AudioQA Case 8]
\textbf{Q.} What type of sound is present in the background of the audio clip?\\
\textit{Options:}\\
A. Car engine \\
B. Ocean waves \\
C. Upbeat synthesized music \\
D. Bird chirping

\vspace{3pt}
\textbf{Internal pred:} B. Ocean waves \\
\textbf{External pred:} C. Upbeat synthesized music

\vspace{3pt}
\begin{qaanswer}
\textbf{Pred:} \texttt{<external>} C. Upbeat synthesized music \\
\textbf{Gold:} \texttt{<external>} C. Upbeat synthesized music
\end{qaanswer}
\end{casecard}

\begin{casecard}[AudioQA Case 9]
\textbf{Q.} Based on the audio description, what is the primary focus of the sounds?\\
\textit{Options:}\\
A. A quiet library setting \\
B. A peaceful nature scene \\
C. A busy city street \\
D. A defense attack scenario

\vspace{3pt}
\textbf{Internal pred:} A. A quiet library setting \\
\textbf{External pred:} D. A defense attack scenario

\vspace{3pt}
\begin{qaanswer}
\textbf{Pred:} \texttt{<external>} D. A defense attack scenario \\
\textbf{Gold:} \texttt{<external>} D. A defense attack scenario
\end{qaanswer}
\end{casecard}

\begin{casecard}[AudioQA Case 10]
\textbf{Q.} Based on the audio description, what type of activity is most likely taking place?\\
\textit{Options:}\\
A. Gardening \\
B. Woodworking \\
C. Lock-picking \\
D. Cooking

\vspace{3pt}
\textbf{Internal pred:} B. Woodworking \\
\textbf{External pred:} C. Lock-picking

\vspace{3pt}
\begin{qaanswer}
\textbf{Pred:} \texttt{<external>} C. Lock-picking \\
\textbf{Gold:} \texttt{<external>} C. Lock-picking
\end{qaanswer}
\end{casecard}

\begin{casecard}[AudioQA Case 11]
\textbf{Q.} What element in the audio contributes to the emotional depth of the song besides the vocals?\\
\textit{Options:}\\
A. The language spoken \\
B. The steady drum beats \\
C. The groovy bass line \\
D. The keyboard harmony

\vspace{3pt}
\textbf{Internal pred:} D. The keyboard harmony \\
\textbf{External pred:} C. The groovy bass line

\vspace{3pt}
\begin{qaanswer}
\textbf{Pred:} \texttt{<internal>} D. The keyboard harmony \\
\textbf{Gold:} \texttt{<internal>} D. The keyboard harmony
\end{qaanswer}
\end{casecard}

\begin{casecard}[AudioQA Case 12]
\textbf{Q.} Based on the audio description, what type of environment is suggested by the background sounds?\\
\textit{Options:}\\
A. A serene forest \\
B. A quiet library \\
C. A beach with waves \\
D. A busy city street

\vspace{3pt}
\textbf{Internal pred:} D. A busy city street \\
\textbf{External pred:} D. A busy city street

\vspace{3pt}
\begin{qaanswer}
\textbf{Pred:} \texttt{<internal>} D. A busy city street \\
\textbf{Gold:} \texttt{<internal>} D. A busy city street
\end{qaanswer}
\end{casecard}

\begin{casecard}[AudioQA Case 13]
\textbf{Q.} What might the purpose of the whistle and crinkling leaves in the background be?\\
\textit{Options:}\\
A. To create a suspenseful atmosphere \\
B. To signal the attention of someone nearby \\
C. To mimic a bustling city environment \\
D. To indicate the presence of wildlife

\vspace{3pt}
\textbf{Internal pred:} B. To signal the attention of someone nearby \\
\textbf{External pred:} B

\vspace{3pt}
\begin{qaanswer}
\textbf{Pred:} \texttt{<internal>} B. To signal the attention of someone nearby \\
\textbf{Gold:} \texttt{<internal>} B. To signal the attention of someone nearby
\end{qaanswer}
\end{casecard}

\begin{casecard}[AudioQA Case 14]
\textbf{Q.} Why does the conversation feature electronic beats and rhythmic cymbal sounds?\\
\textit{Options:}\\
A. To mimic the sounds of a busy environment \\
B. To create a sense of urgency in the interaction \\
C. To drown out background noise \\
D. To enhance the emotional depth of the speech

\vspace{3pt}
\textbf{Internal pred:} B. To create a sense of urgency in the interaction \\
\textbf{External pred:} D. To enhance the emotional depth of the speech

\vspace{3pt}
\begin{qaanswer}
\textbf{Pred:} \texttt{<internal>} B. To create a sense of urgency in the interaction \\
\textbf{Gold:} \texttt{<internal>} B. To create a sense of urgency in the interaction
\end{qaanswer}
\end{casecard}

\begin{casecard}[AudioQA Case 15]
\textbf{Q.} What sound appears earliest in the audio?\\
\textit{Options:}\\
A. Ticking \\
B. Accelerating, revving, vroom \\
C. Idling \\
D. Car

\vspace{3pt}
\textbf{Internal pred:} C. Idling \\
\textbf{External pred:} D. Car

\vspace{3pt}
\begin{qaanswer}
\textbf{Pred:} \texttt{<internal>} C. Idling \\
\textbf{Gold:} \texttt{<internal>} C. Idling
\end{qaanswer}
\end{casecard}

\end{document}